\documentclass[
aps,
pre,
amsmath,amssymb,amsfonts,
reprint,
floatfix,
nofootinbib,
]{revtex4-2}

\usepackage{graphicx}
\usepackage{dcolumn}
\usepackage{bm}

\usepackage{mathptmx}
\usepackage{amscd}
\usepackage{xcolor}

\usepackage[caption=false]{subfig}
\usepackage{ragged2e}
\DeclareCaptionJustification{justified}{\justifying}

\usepackage{anyfontsize}

\usepackage{amsthm}

\newtheoremstyle{my_theorem_style_remark}{2.5pt}{2.5pt}{}{}{\slshape\bfseries}{.}{  }{\thmname{#1}\thmnumber{ #2}}
\theoremstyle{my_theorem_style_remark}
\newtheorem{remark}{Remark}
\newtheoremstyle{my_theorem_style}{2.5pt}{2.5pt}{\itshape}{}{\bfseries}{.}{  }{}
\theoremstyle{my_theorem_style}
\newtheorem{proposition}{Proposition}
\newtheorem{corollary}{Corollary}
\newtheorem{lemma}{Lemma}

\allowdisplaybreaks

\begin{document}

\title{Many interacting particles in solution. III. Spectral analysis of the associated Neumann--Poincar\'e-type operators}
\author{Sergii V. Siryk}
\email{accandar@gmail.com}
\affiliation{CONCEPT Lab, Istituto Italiano di Tecnologia, Via E. Melen 83, 16152, Genova, Italy}
\author{Walter Rocchia}
\email[W. Rocchia \emph{(corresponding co-author)}: ]{walter.rocchia@iit.it}
\affiliation{CONCEPT Lab, Istituto Italiano di Tecnologia, Via E. Melen 83, 16152, Genova, Italy}

\begin{abstract}
\noindent
The interaction of particles in an electrolytic medium can be calculated by solving the Poisson equation inside the solutes and the linearized Poisson--Boltzmann equation in the solvent, with suitable boundary conditions at the interfaces. Analytical approaches often expand the potentials in spherical harmonics, relating interior and exterior coefficients and eliminating some coefficients in favor of others, but a rigorous spectral analysis of the corresponding formulations is still lacking. Here, we introduce pertinent composite many-body Neumann--Poincar\'e-type operators and prove that they are compact with spectral radii strictly less than one. These results provide the foundation for systematic screening-ranged expansions, in powers of the Debye screening parameters, of electrostatic potentials, interaction energies, and forces, and establish the analytical framework for the accompanying works~\textcolor{red}{\cite{supplem_prl,supplem_pre,supplem_pre_force}}.
\end{abstract}

\maketitle
\section{Introduction}
\label{section-Intro}
\noindent
The Poisson-Boltzmann equation (PBE) is a key tool in electrostatic modeling of soft matter, plasma physics, colloid science, and biophysics. In colloid sciences, it is widely used to understand self-assembly processes and to design new materials with specific electrostatic properties \cite{LGDLK,BarrLui_2014,LDKCSB}, while in biophysical applications it is used to calculate electrostatic potentials and estimate the electrostatic energy of biomolecular systems immersed in electrolytic solutions. It is often adopted in drug design studies and for modeling interactions at the molecular level (such as protein–protein and protein-ligand binding, nucleic acid stability, enzyme activity, etc.)~\cite{JPCBSIYR,Blossey2023,BesleyACR}. In general, the PBE is a second-order elliptic nonlinear partial differential equation, the solution of which poses challenges not only analytically but also numerically~\cite{DiFlorio2025NextGenPB}. In systems generating low potential values—a case of great practical interest—the PBE can be linearized, yielding the simpler linearized PBE (LPBE), or Debye-H\"uckel (DH) equation. This linear form is more manageable both numerically and analytically than its nonlinear counterpart. Nonetheless, even for highly charged solutes, the LPBE remains insightful (especially at distances large compared to the Debye length), provided the electric field sources are properly renormalized \cite{Alex1,TrizPRL,Triz1,ST1,Boon2015,our_jcp,our_jpcb,Krishnan2017,SchlaichHolm,BritoDenton,BoonJanus2010}. These considerations make the LPBE the most commonly used equation for studying the electrostatics of biomolecular systems. From the standpoint of rigorous analytical treatment, even the seemingly simple case of two interacting dielectric spheres immersed in a solvent leads to intricate mathematical analyses (recent reviews of this longstanding problem are available in~\cite{BesleyACR,our_jcp,our_jpcb}).

The search for self-consistent analytical or semi-analytical solutions to the LPBE for spherical solutes (systems composed of spherical solutes are of long-standing interest due to their mathematical tractability and physical relevance \cite{our_jcp,our_jpcb,Derb2,Derb4,BesleyACR,BSL2014,BarrLui_2014,Yu2019,Yu3,Yu2021,WL_prl_2019,Qin2019,CurLui_2021}) typically involves expanding both internal and external potentials into spherical harmonics eigen-series. These are then coupled via boundary/transmission conditions at the media interfaces (i.e.~the solutes' surfaces). The resulting system of linear equations, originally of infinite size (since the number of terms and corresponding multipolar coefficients in the series is infinite), is usually truncated at a certain degree and solved numerically. Analytical solutions are rare and often require severe simplifying assumptions \cite{our_jcp,Derb2,Derb4,Fish}, leading to potentially rough approximations of the true LPBE solution. Interestingly, approaches for exact LPBE solutions in two-sphere systems without such assumptions have recently been proposed in~\cite{our_jcp}. A commonly adopted approach to solve these systems (see e.g.~\cite{Derb2,Derb4,GWJXL,our_jcp,our_jpcb,LianMa,BP1,Head-Gordon2006,Fish}) consists of eliminating the internal expansion coefficients and deriving equations that involve only the external coefficients; once the latter are known, the internal coefficients follow by back-substitution. This significantly reduces the number of unknowns and, as shown in recent work (see \cite{our_jcp} and also \textcolor{red}{\cite{supplem_prl,supplem_pre,supplem_pre_force}}), forms the basis for further developments in both analytical theory (e.g.~constructing LPBE-rigorous expressions quantifying interactions beyond pairwise DLVO-type (Derjaguin-Landau-Verwey-Overbeek) approximations) and fast numerical approaches. This procedure, however, comes at the cost of leading to a significantly more complicated form of the resulting system of equations and hinders the realization of a rigorous study of its properties -- particularly, the spectral properties \cite{BSL2014,WuLui_jcp_18} of the corresponding matrices/operators, whether in conventional truncated finite-dimensional approximations or in the infinite-dimensional setting. The latter analysis, though quite subtle, is essential for constructing the exact explicit screening-ranged series solution to the LPBE, with guaranteed convergence, elaborated in \textcolor{red}{\cite{supplem_prl,supplem_pre}} (see Remark~\ref{remark_operator_neumann_series} below). This paper seeks to fill this gap by providing, to our knowledge for the first time, a rigorous infinite-dimensional spectral analysis of the many-body problem of dielectric spheres within the LPBE framework. 

To this end, we proceed as follows:
\newcounter{my_temp_list_counter0}
\begin{list}{{\arabic{my_temp_list_counter0}.} }{\usecounter{my_temp_list_counter0} \topsep=0pt \parsep=0pt \itemsep=0pt \partopsep=0pt \parskip=0pt \itemindent=0pt \leftmargin=0pt \labelwidth=0pt \labelsep=0pt}
    \item\label{nov_list_exact_sol}We begin by stating the boundary-value problem for the self-consistent total (i.e.~valid inside and outside the solutes) electrostatic potential (Sec.~\ref{basic_problem_statement0}). We then reformulate it into specific Boundary-Integral-Equation-type formulations involving either the surface DH potentials or the corresponding ``effective'' surface charge densities (Sec.~\ref{basic_BIE_formulations}).
    \item\label{nov_list_connect}We then establish a correspondence between these continuous formulations and a discrete system that couples the spherical harmonics expansion coefficients of external (DH) potentials (Sec.~\ref{spherical_Fourier_coeffs_subsec}). While this system is largely out of reach for conventional discrete analysis approaches such as infinite-dimensional matrix analysis methods, the above-mentioned correspondence allows the application of functional-analytic and abstract potential theory arguments.
    \item\label{nov_list_spectra}Finally, we rigorously study the spectral properties of the associated infinite-dimensional operators arising in both the continuous and discrete settings (Sec.~\ref{spectral_properties_subsection}). We prove that the global many-body block Neumann--Poincar\'e-type (NP-type) operators\footnote{See definitions of operators $\mathcal K$, $\mathring{\mathcal K}$ and $\mathbb K$ in Sec.~\ref{statements_basics_section}.} are compact—ensuring ``discreteness"\footnote{See Sec.~\ref{appendix_compact_fredholm_operators_summary} for precise definitions.} of spectra—and have spectral radii strictly less than one.
\end{list}

Although this work is primarily aimed at rigorously underpinning the analytical formalism of screening-ranged expansions of electrostatic quantities developed in \textcolor{red}{\cite{supplem_prl,supplem_pre,supplem_pre_force}}, we believe it has independent theoretical value. Indeed, while the study of multi-domain/multiple-boundary/many-body NP-type operators is a rapidly growing field in modern mathematical physics (see for instance \cite{Ammari_Ciraolo_1,Kang_Kim_1,CLS_siam,AMRZ_2017,LLW_2025,AKMP_2025,DKPZZ,JiKang2023}), the spectral analysis of the specific formulations studied in this work appears to be novel. To guarantee mathematical rigor, we provide all proofs in Sec.~\ref{Proofs_appendix_all_proposition}, relying on recent advances in functional analysis and operator theory \cite{Ammari_Ciraolo_1,Kang_Kim_1}, fine estimates of modified Bessel functions \cite{Segura_jmaa_2011}, Sobolev space embedding theorems \cite{SauterSchwab}, and properties of Wigner-rotation-free spherical re-expansions advanced in recent works \cite{Yu3,Yu2021}, the last being crucial in the construction of the discrete NP-type operators.

\section{Boundary value problems and equivalent Boundary-Integral-Equations formulations}
\label{statements_basics_section}
\subsection{Problem statement}
\label{basic_problem_statement0}
\noindent
Let us consider a general system consisting of $N$ non-overlapping spherical dielectric particles represented by balls $\Omega_i$, $i\in\{1,\ldots,N\}$ (or, in a simpler notation and to emphasize the running character of index $i$, $i=\overline{1,\ldots,N}$), where $\Omega_i\subset\mathbb R^3$ is an open set in $\mathbb R^3$. These particles are immersed into the surrounding medium (electrolytic solvent -- e.g.~water and mobile ions) described by dielectric constant $\varepsilon_\text{sol}$ and Debye screening length $\kappa^{-1}>0$. Let us also note that since in this study $\kappa$ is treated rather as a parameter (of the LPBE \eqref{Lin_eqs_lpb} -- see below), its precise definition \cite{LBD2023,Kjellander_JCP_2016} is therefore not binding to us. Each particle $\Omega_i$ is centered at $\mathbf x_i\in\mathbb R^3$ and is characterized by its dielectric constant $\varepsilon_i$ and radius~$a_i$. The electrostatic potential $\Phi_{\text{in},i}(\mathbf r)$ in $\Omega_i$ (i.e., as $r_i < a_i$ where $r_i$ is the radial coordinate of $\mathbf r \in \mathbb{R}^3$ measured from $\mathbf x_i$ so that $r_i = \left\|\mathbf r_i\right\|$, $\mathbf r_i = \mathbf r - \mathbf x_i$) satisfies the Poisson equation (PE), while the corresponding potential $\Phi_{\text{out},i}$ in solvent, due to the presence of the $i$-th particle, fulfills the LPBE~\cite{our_jcp,our_jpcb}:
\begin{subequations}
\label{Lin_eqs}
\begin{align}
&\Delta\Phi_{\text{in},i}(\mathbf r)= - \rho_i^\text{f}(\mathbf r)/(\varepsilon_i \varepsilon_0), & & \mathbf r\in\Omega_i,\label{Lin_eqs_poisson} \\
&\Delta\Phi_{\text{out},i}(\mathbf r) - \kappa^2 \Phi_{\text{out},i}(\mathbf r) = 0, & & \mathbf r\in\Omega_\text{sol}\label{Lin_eqs_lpb}
\end{align}
\end{subequations}
for all $i=\overline{1,\ldots,N}$, where $\rho_i^\text{f}(\mathbf r)$ denotes the free (fixed) charge density distribution supported inside the $i$-th particle, and the solvent domain $\Omega_\text{sol} \mathrel{:=} \mathbb R^3\setminus\bigcup_{i=1}^N\overline{\Omega}_i$ (where $\overline{\Omega}_i$ is the $\mathbb R^3$-closure of $\Omega_i$). Due to the superposition principle adopted in the DH description \cite{Fish,Derb2} the total self-consistent electrostatic potential $\Phi(\mathbf r)$ of the whole system is~then
\begin{equation}
\label{Lin_eq_tot_pot}
\Phi(\mathbf r) = \left[
\begin{aligned}
&\Phi_{\text{in},i}(\mathbf r),\quad \mathbf r\in\Omega_i,\\
&\Phi_{\text{out}}(\mathbf r) \mathrel{:=} \sum\nolimits_{i=1}^N \Phi_{\text{out},i}(\mathbf r),\quad \mathbf r\in\Omega_\text{sol},
\end{aligned}
\right.
\end{equation} 
while at the boundaries between different media the potential $\Phi(\mathbf r)$ is subject to transmission type boundary conditions~(BCs)
\begin{subequations}
\label{Lin_eq_standard_bc}
\begin{gather}
\left.\Phi_{\text{in},i}\right|_{r_i \to a_i^-} = \left.\Phi_{\text{out}}\right|_{r_i \to a_i^+}, \label{Lin_eq_standard_bc_1st} \\
\varepsilon_i\left.(\mathbf n_i\cdot\nabla\Phi_{\text{in},i})\right|_{r_i\to a_i^-} - \varepsilon_\text{sol}\left.(\mathbf n_i\cdot\nabla\Phi_{\text{out}})\right|_{r_i\to a_i^+} = \sigma_i^\text{f}/\varepsilon_0 \label{Lin_eq_standard_bc_2nd}
\end{gather}
\end{subequations}
for all $i=\overline{1,\ldots,N}$, where $\mathbf n_i$ is the outer unit normal and $\sigma_i^\text{f}$ is a free charge density (if any) on the boundary $\partial\Omega_i$ ($r_i=a_i$) of $\Omega_i$, and $r_i\to a_i^{\pm}$ denotes approaching $\partial\Omega_i$ from exterior($+$)/interior($-$) of that particle. For the definiteness, we will focus on the transmission BCs of type \eqref{Lin_eq_standard_bc} which are of primary interest for biomolecular sciences \textcolor{red}{\cite{supplem_prl,supplem_pre}}. Extensions of the proposed analysis to other types of BCs (fixed potentials, linear charge regulation BCs) and incorporation of the Stern layer into the problem statement are deferred to future work. Further, we assume that all $\sigma_i^\text{f}=0$ in \eqref{Lin_eq_standard_bc_2nd}. This does not bring any loss of generality, since both problem formulations are mathematically similar in essence (as we will see below, this modification affects only right-hand sides of the resulting system of equations governing DH potentials); moreover, the case of $\sigma_i^\text{f}\ne0$ will be incorporated later as well (see Remark~\ref{free_surf_charge_remark}).

The potential $\Phi_{\text{in},i}$ can be decomposed \cite{our_jcp,our_jpcb}~as $$\Phi_{\text{in},i} = \Hat\varPhi_{\text{in},i} + \Tilde\Phi_{\text{in},i} ,$$ where $\Hat\varPhi_{\text{in},i}(\mathbf r) = \frac{1}{4\pi\varepsilon_0\varepsilon_i}\!\int_{\Omega_i}\!\frac{\rho_i^\text{f}(\mathbf r') \, d \mathbf r'}{\|\mathbf r - \mathbf r' \|}$ is the particular solution to \eqref{Lin_eqs_poisson} representing the standard Coulombic potential, screened by local polarization, in infinite space produced by $\rho_i^\text{f}$, while $\Tilde\Phi_{\text{in},i}$ fulfills the Laplace equation, $\Delta\Tilde\Phi_{\text{in},i}=0$. We also remind the conventional conditions ensuring physical feasibility~\cite{our_jcp,Jack}, namely $|\Tilde\Phi_{\text{in},i}|<\infty$ as $r_i\to0^+$ and $\Phi_{\text{out},i}\to0$ as $r_i\to+\infty$.

\subsection{Layered potential operators and general boundary-integral-equations (BIE) formulations}
\label{basic_BIE_formulations}
\noindent
Let us look for unknown potentials $\Tilde\Phi_{\text{in},i}$ and $\Phi_{\text{out},i}$ in the form~\cite{Head-Gordon2010,BordHub2003,Chipman2004,BardhanJCP2009}
\begin{equation}
\label{vol_potentials_via_operators}
\Tilde\Phi_{\text{in},i} = \Breve{\mathcal S}_i^0 \mathfrak l_i, \qquad
\Phi_{\text{out},i} = \Breve{\mathcal S}_i^\kappa \mathfrak g_i
\end{equation}
with superficial (yet to be determined) polarization charge density $\mathfrak l_i$ and effective charge density $\mathfrak g_i$ both belonging to $L^2(\partial\Omega_i)$ (where $L^2$ denotes spaces of square-integrable Lebesgue-measurable functions), where 
\begin{equation}
\label{S_vol_definition}
\Breve{\mathcal S}_i^\kappa\phi (\mathbf r) \mathrel{:=} - \oint_{\partial\Omega_i}\frac{e^{-\kappa \|\mathbf r-\mathbf s\|}}{4\pi \|\mathbf r-\mathbf s\|} \phi(\mathbf s) d s ,\qquad \mathbf r\in\mathbb R^3
\end{equation}
(for $\mathbf r\in \partial\Omega_i$ it is defined as an improper integral) is a screened volumetric single-layer potential operator \cite{Head-Gordon2010,BordHub2003,BardhanJCP2009,Chipman2004,AmmariKang_mathstat} (here $d s$ is a surface measure on $\partial\Omega_i$). Its limiting restriction to the surface $\partial\Omega_i$ itself (i.e.~the corresponding boundary single-layer operator on $\partial\Omega_i$) we denote by $\mathcal S_i^\kappa$ -- more precisely, $\mathcal S_i^\kappa \mathrel{:=} \gamma_{\partial\Omega_i} \Breve{\mathcal S}_i^\kappa$, where $\gamma_{\partial\Omega_i}$ is the corresponding trace (restriction on the boundary $\partial\Omega_i$) operator, see Remark~\ref{remark_the_same_notation_potentials_surface_volumetric} for technical details. From the (well-known in potential theory, see e.g.~\cite[Chap.~2]{AmmariKang_mathstat}) jump relations for the traces of the derivatives of layer potentials we then obtain for the normal derivatives $\frac{\partial}{\partial\mathbf n_i^\pm}$ on $\partial\Omega_i$ (superscript $\pm$ means approaching $r_i\to a_i^\pm$ so that $\frac{\partial}{\partial\mathbf n_i^\pm}(\cdot) \mathrel{:=} \frac{\partial}{\partial\mathbf n_i}(\cdot)\bigr|_{r_i\to a_i^\pm}$) in our notation: $\frac{\partial}{\partial\mathbf n_i^\pm}\Breve{\mathcal S}_i^\kappa\phi = \bigl((\pm 1/2)\mathcal I_i + \mathcal K_i^\kappa{}^\star \bigr)\phi$, where $\mathcal I_i$ is the identity operator and $\mathcal K_i^\kappa{}^\star$ is the (adjoint) elementary ``one-body" NP operator \cite{AmmariKang_mathstat,SauterSchwab,AKMP_2025} on $\partial\Omega_i$, $$\mathcal K_i^\kappa{}^\star \phi(\mathbf r) \mathrel{:=} - \oint_{\partial\Omega_i}\mathbf n_i\cdot\nabla_{\mathbf r} \, \frac{e^{-\kappa \|\mathbf r-\mathbf s\|}}{4\pi \|\mathbf r-\mathbf s\|} \phi(\mathbf s) d s .$$ 
Thus, BC~\eqref{Lin_eq_standard_bc_1st} yields $$\mathcal S_i^0 \mathfrak l_i (\mathbf r) + \Hat\varPhi_{\text{in},i}(\mathbf r)|_{r_i\to a_i^-} = \mathcal S_i^\kappa \mathfrak g_i(\mathbf r) + \sum_{j=1,\,j\ne i}^N \gamma_{\partial\Omega_i} \Breve{\mathcal S}_j^\kappa \mathfrak g_j(\mathbf r)$$ and BC~\eqref{Lin_eq_standard_bc_2nd} yields (by employing the above jump relations for expressing $\frac{\partial\mathcal S_i^\kappa\phi}{\partial\mathbf n_i^\pm}$)
\begin{align*}
& \varepsilon_i\Bigl(-\frac{1}{2}\mathcal I_i + \mathcal K_i^0{}^\star\Bigr) \mathfrak l_i(\mathbf r) = \varepsilon_\text{sol}\Bigl(\frac{1}{2}\mathcal I_i + \mathcal K_i^\kappa{}^\star\Bigr) \mathfrak g_i(\mathbf r) \\
&\quad + \varepsilon_\text{sol}\sum_{j=1,\,j\ne i}^N \frac{\partial}{\partial\mathbf n_i^+}\Breve{\mathcal S}_j^\kappa \mathfrak g_j(\mathbf r) -\varepsilon_i \frac{\partial}{\partial\mathbf n_i^-}\Hat\varPhi_{\text{in},i}(\mathbf r) 
\end{align*}
for arbitrary $\mathbf r\in\partial\Omega_i$. Now, by eliminating the density $\mathfrak l_i$ by using the first relation in the second one (it is permitted since the inverse $(\mathcal S_i^0 )^{-1}$ is correctly defined here -- see Remark~\ref{remark_single_layer_potential_mapping_property} below) we then arrive at a system of operator equations
\begin{equation}
\label{main_eq_G_op}
\mathcal A_i \mathfrak g_i (\mathbf r) + \sum_{j=1,\, j\ne i}^N \mathcal B_{i j} \mathfrak g_j (\mathbf r) = \mathfrak s_i(\mathbf r), \qquad \mathbf r\in\partial\Omega_i,
\end{equation}
where the operators $\mathcal A_i$, $\mathcal B_{i j}$, and the right-hand sides $\mathfrak s_i$ are:
\begin{align*}
& \mathcal A_i \mathrel{:=} \varepsilon_i\Bigl( -\frac{1}{2}\mathcal I_i + \mathcal K_i^0{}^\star \Bigr)(\mathcal S_i^0 )^{-1}\mathcal S_i^\kappa - \varepsilon_\text{sol}\Bigl( \frac{1}{2}\mathcal I_i + \mathcal K_i^\kappa{}^\star \Bigr)\!, \\
& \mathcal B_{i j} \mathrel{:=} \varepsilon_i\Bigl(-\frac{1}{2}\mathcal I_i + \mathcal K_i^0{}^\star \Bigr)(\mathcal S_i^0 )^{-1} \gamma_{\partial\Omega_i} \Breve{\mathcal S}_j^\kappa  - \varepsilon_\text{sol} \frac{\partial}{\partial\mathbf n_i^+}\Breve{\mathcal S}_j^\kappa , \\
& \mathfrak s_i \mathrel{:=} \varepsilon_i\Bigl(-\frac{1}{2}\mathcal I_i + \mathcal K_i^0{}^\star \Bigr)(\mathcal S_i^0 )^{-1}\Hat\varPhi_{\text{in},i}|_{r_i\to a_i^-} - \varepsilon_i\frac{\partial}{\partial\mathbf n_i^-}\Hat\varPhi_{\text{in},i} 
\end{align*}
$\forall i\in\overline{1,\ldots,N}$. One thus sees that the operator formulation of the system of equations coupling the external densities $\{\mathfrak g_i\}_{i=1}^N$ and obtained by completely eliminating the internal densities $\{\mathfrak l_i\}_{i=1}^N$ acquires a rather complex structure; to the best of our knowledge, spectral analysis of such a system has never been carried out before (in general, not to mention infinite-dimensional spectral analyses of multi-region dielectric problems, even the spectral properties of truncated discretized single-object formulations can rarely be studied analytically~\cite{BSL2014,WuLui_jcp_18}). 

Let us introduce the direct sum $\mathbf L^2 \mathrel{:=} \bigoplus\nolimits_{i=1}^N L^2(\partial\Omega_i)$ and composite block $\mathbf L^2\to\mathbf L^2$ operators  $\mathcal A \mathrel{:=} \operatorname{diagonal}\{\mathcal A_i\}_{i=1}^N$, $\mathcal B \mathrel{:=} \{\mathcal B_{i j}\}_{i, j=1; j\ne i}^N$, $\mathcal K \mathrel{:=} \mathcal A^{-1}\mathcal B = \{\mathcal K_{i j}\}_{i, j=1; j\ne i}^N$ (where $\mathcal K_{i j} \mathrel{:=} \mathcal A_i^{-1}\mathcal B_{i j}$, while diagonal components $\mathcal K_{i i}$ are formally put as zero operators) -- the expanded form of operator $\mathcal K$~is:
\begin{gather*}
\mathcal K =
\begin{pmatrix}
0 & \mathcal A_1^{-1} \mathcal B_{1 2} & \ldots & \mathcal A_1^{-1} \mathcal B_{1 N} \\
\mathcal A_2^{-1} \mathcal B_{2 1} & 0 & \ldots & \mathcal A_2^{-1} \mathcal B_{2 N} \\
\vdots & \vdots & \ddots & \vdots \\
\mathcal A_N^{-1} \mathcal B_{N 1} & \mathcal A_N^{-1}\mathsf B_{N 2} & \ldots & 0
\end{pmatrix}\!. 
\end{gather*} 
It will be proven further (see Lemma~\ref{Ai_has_bounded_inverse} in Sec.~\ref{Proof_appendix_compactness_proposition}) that every operator $\mathcal A_{i} \colon L^2(\partial\Omega_i)\to L^2(\partial\Omega_i)$ is bounded and has a bounded inverse, thus $\mathcal A^{-1}$ and hereby $\mathcal K$ are correctly defined here. System \eqref{main_eq_G_op} then reads in block matrix-operator form as $(\mathcal A+\mathcal B)\Vec{\mathfrak g} = \Vec{\mathfrak s}$, or equivalently 
\begin{equation}
\label{main_eq_G_op_L2}
(\mathcal I + \mathcal K)\Vec{\mathfrak g} = \mathcal A^{-1}\Vec{\mathfrak s},
\end{equation}
where column-vectors $\Vec{\mathfrak g} \mathrel{:=} \{\mathfrak g_i\}_{i=1}^N$ and $\Vec{\mathfrak s} \mathrel{:=} \{\mathfrak s_i\}_{i=1}^N$ belong to $\mathbf L^2$, and $\mathcal I$ is a generic designation for an identity operator acting on spaces under consideration. 

Finally, introducing $\mathbf H^s \mathrel{:=} \bigoplus\nolimits_{i=1}^N H^s(\partial\Omega_i)$ ($H^s$ denotes Sobolev spaces, see Remark~\ref{remark_the_same_notation_potentials_surface_volumetric}), acting on both sides of \eqref{main_eq_G_op_L2} by composite block operator $\mathcal S^\kappa \mathrel{:=} \operatorname{diagonal}\{\mathcal S_i^\kappa\}_{i=1}^N \colon \mathbf L^2 \to \mathbf H^1$ (see Remark~\ref{remark_single_layer_potential_mapping_property}), and taking into account that $\Vec{\mathfrak g} = (\mathcal S^\kappa)^{-1}\Vec{\Phi}_{\text{out}}^{\partial\Omega}$, where the column-vector of boundary potentials $\Vec{\Phi}_{\text{out}}^{\partial\Omega} \mathrel{:=} \{\left.\Phi_{\text{out},i}\right|_{\partial\Omega_i}\}_{i=1}^N \in \mathbf H^1$, we can then reduce \eqref{main_eq_G_op_L2} to the following equivalent form (but now acting on $\mathbf H^1$ instead of \eqref{main_eq_G_op_L2} which acts on~$\mathbf L^2$):
\begin{equation}
\label{main_eq_G_op_Sobolev}
(\mathcal I + \mathring{\mathcal K}) \Vec{\Phi}_{\text{out}}^{\partial\Omega} = \mathcal S^\kappa \mathcal A^{-1}\Vec{\mathfrak s},
\end{equation}
where block operator $\mathring{\mathcal K} \mathrel{:=} \mathcal S^\kappa \mathcal K (\mathcal S^\kappa)^{-1}$. Proposition~\ref{proposition_matrix_repr} below elucidates the benefit of introducing formulation~\eqref{main_eq_G_op_Sobolev}. 

By an analogy with the Fredholm integral equations corresponding to the classical ``one-body" boundary-value problems \cite{NedelecAEE,AmmariKang_mathstat,SauterSchwab, McCamyStephan} where the simplest/elementary NP operator $\mathcal K_i^\kappa{}^\star$ plays the role of a perturbation to some invertible operator (e.g.~the identity operator), and following the terminology quite established in the recent mathematical literature \cite{Kang_Kim_1,Ammari_Ciraolo_1} for operators of such type, we will also call the global many-body composite/block operators $\mathcal K$ and $\mathring{\mathcal K}$ of \eqref{main_eq_G_op_L2}-\eqref{main_eq_G_op_Sobolev} the \emph{NP-type} operators.

The formulations \eqref{main_eq_G_op_L2} and \eqref{main_eq_G_op_Sobolev} are quite general and can be applied also to systems of non-spherical dielectric particles (the sphericity of $\partial\Omega_i$ did not play any explicit role in their derivation).

Let us also note that in the majority of works aimed at solving \eqref{Lin_eqs} for spherical particles (see overviews in \textcolor{red}{\cite{supplem_prl,supplem_pre,supplem_pre_force}} and references therein) a different and more conventional approach is pursued, namely consisting in expansion potentials $\Tilde\Phi_{\text{in},i}(\mathbf r)$ and $\Phi_{\text{out},i}(\mathbf r)$ through eigenfunctions Fourier-type series ($\{r_i Y_n^m (\Hat{\mathbf r}_i)\}$ for PE, $\{k_n(\kappa r_i) Y_n^m(\Hat{\mathbf r}_i)\}$ for LPBE, see \eqref{Lin_eqs_Phi_in_out} below) with some unknown potential coefficients to be determined by using BCs. Following such an approach (see Sec.~\ref{spherical_Fourier_coeffs_subsec} below), however, we have realized that it is extremely challenging (see further technical comments in Sec.~\ref{spherical_Fourier_coeffs_subsec}, \ref{spectral_properties_subsection}), if not impossible, to directly investigate the spectrum and properties of the corresponding original ``discrete" global NP-type operator (operator $\mathbb K$ -- see \eqref{global_lin_sys1} below) emerging in this situation and acting on the square-summable sequences of potential coefficients. Therefore, we circumvent this problem by studying operators $\mathcal K$ and $\mathring{\mathcal K}$ instead (using the methods of modern abstract potential theory, functional analysis and PDEs) and then deriving the corresponding properties for~$\mathbb K$.
\begin{remark}
\label{different_BIE_formulations_remark}
There are various (essentially equivalent) ways of representing the solution to the coupled PE-LPBE problem and the corresponding BIE formulations \cite{BordHub2003,Chipman2004,BardhanJCP2009,OMHOOC,Altm_Bard_JCC_,DiFlorio2025NextGenPB,HildBlos_PRL}. The formulation of Bordner and Huber \cite{BordHub2003}, which uses two different single-layer densities for representing the solute's (obeying PE) and the solvent's (obeying LPBE) potentials, is one of the most popular and widely quoted in the literature on numerical solutions (especially using BEM-type approaches) to this coupled problem \cite{Head-Gordon2010,BordHub2003,Chipman2004,BardhanJCP2009}, though a rigorous spectral analysis of this formulation has to date remained elusive. Representation \eqref{vol_potentials_via_operators} invoking pairs $(\mathfrak l_i,\mathfrak g_i)$ of single-layer distributions can thus be considered in this sense as some kind of a many-body generalization of the Bordner-Huber formulation with excluded internal densities.
\end{remark}

\subsection{Representation in the spherical Fourier coefficients space}
\label{spherical_Fourier_coeffs_subsec}
\noindent
Expanding the (rigid-motions invariant) integral kernels $\frac{1}{4\pi \|\mathbf r_i-\mathbf s_i\|}$ and $\frac{e^{-\kappa \|\mathbf r_i-\mathbf s_i\|}}{4\pi \|\mathbf r_i-\mathbf s_i\|}$ of $\Breve{\mathcal S}_i^0$ and $\Breve{\mathcal S}_i^\kappa$ in \eqref{vol_potentials_via_operators} through expansions for the corresponding Green functions in spherical coordinates (see \eqref{addition_theorem_screened_0}-\eqref{Phi_i_expansion_addition_t} below) with harmonics centered at $\mathbf x_i$ we immediately arrive at relations 
\begin{subequations}
\label{Lin_eqs_Phi_in_out}
\begin{gather}
\Tilde\Phi_{\text{in},i}(\mathbf r) = \sum\nolimits_{n,m}L_{n m,i} \Tilde r_i^n Y_n^m(\Hat{\mathbf r}_i),\label{Lin_eqs_Phi_in} \\
\Phi_{\text{out},i}(\mathbf r) = \sum\nolimits_{n,m}G_{n m,i}k_n(\Tilde r_i) Y_n^m(\Hat{\mathbf r}_i)\label{Lin_eqs_Phi_out}
\end{gather}
\end{subequations}
with coefficients 
\begin{equation}
\label{expansion_surface_densities_coeffs}
\begin{aligned}
L_{n m,i} & = \frac{-\kappa}{(2 n+1)\Tilde a_i^{n+1}} \oint_{\partial\Omega_i} \mathfrak l_i(\Hat{\mathbf s}_i) Y_n^m(\Hat{\mathbf s}_i)^\star d s_i,\\
G_{n m,i} & = -\kappa i_n(\Tilde a_i) \oint_{\partial\Omega_i} \mathfrak g_i(\Hat{\mathbf s}_i) Y_n^m(\Hat{\mathbf s}_i)^\star d s_i 
\end{aligned}
\end{equation}
(superscript $\star$ here stands for the complex conjugation) related to charge densities $\mathfrak l_i$, $\mathfrak g_i$. Here $i=\overline{1,\ldots,N}$, the scaled dimensionless radial variable $\Tilde r_i \mathrel{:=} \kappa r_i$, $\Tilde a_i \mathrel{:=} \kappa a_i$, the sum $\sum_{n,m} \mathrel{:=} \sum_{0\le|m|\le n} =  \sum_{n=0}^{+\infty} \sum_{m=-n}^n$ (also notation $0\le\left|m\right|\le n$ for indices $n$ and $m$ henceforth means the running $n=\overline{0,\ldots,+\infty}$, $m=\overline{-n,\ldots,n}$), and unit vector $\Hat{\mathbf r}_i \mathrel{:=} \mathbf r_i/r_i$ (spherical angles $\theta_i$ and $\varphi_i$ emanating from $\Hat{\mathbf r}_i$ and then used in (complex-valued) spherical harmonics~\eqref{Ynm_definition}, are measured for any $i$ relative to a local coordinate system with center at $\mathbf x_i$ and axes parallel to those of some unique (fixed) global coordinate system); $k_n(x) \mathrel{:=} \sqrt{2/\pi}K_{n+1/2}(x)/\sqrt{x}$ and $i_n(x) \mathrel{:=} \sqrt{\pi/2}I_{n+1/2}(x)/\sqrt{x}$ are modified spherical Bessel functions of the 2nd and 1st kind (see Appendix~\ref{appendix_bessel_functions_summary}). The unknown coefficients $\{L_{n m,i}\}$ and $\{G_{n m,i}\}$ are to be determined from BCs. Next, for a given $\rho_i^\text{f}$ (supported inside $\Omega_i$) potential $\Hat\varPhi_{\text{in},i}$ can also be expanded around $r_i \to a_i^-$ in multipoles 
\begin{equation}
\label{varPhi_in_i_multipoles}
\Hat\varPhi_{\text{in},i} (\mathbf r) = \sum\nolimits_{n,m} \Hat L_{n m,i} \Tilde r_i^{-n-1} Y_n^m(\Hat{\mathbf r}_i)
\end{equation}
with spherical multipole moments (see \cite[Eqs.~(11)-(12)]{our_jcp}) 
\begin{equation}
\label{varPhi_in_i_multipoles_Hat_Lmn}
\Hat L_{n m,i} = \frac{\kappa}{(2 n+1)\varepsilon_i \varepsilon_0}\int_{\Omega_i} \rho_i^\text{f}(\mathbf r_i) \Tilde r_i^n Y_n^m(\Hat{\mathbf r}_i)^\star d\mathbf r_i .
\end{equation}
Now using \eqref{Lin_eqs_Phi_in_out}, \eqref{varPhi_in_i_multipoles}, and employing re-expansion\footnote{Note that the proof of re-expansions \eqref{Yu3_reexp}-\eqref{coeffs_HlmLM_definition}, given in \cite[Appendix~A]{Yu3}, requires the additional condition $r_j>R_{i j}$ is also met (which may formally prevent from applying \eqref{Yu3_reexp}-\eqref{coeffs_HlmLM_definition} on the full spherical surface). For the sake of completeness, in Appendix~\ref{Yu_reexpansion_restriction_removing} below we gain further insight into the nature of these re-expansions and briefly discuss a simple analytical argumentation justifying that the condition $r_i<R_{i j}$ alone is sufficient for \eqref{Yu3_reexp}-\eqref{coeffs_HlmLM_definition} to hold; note also that in our study the last inequality is always met since we actually employ re-expansions on particles' boundaries ($r_i=a_i$).\label{footnote_Yu3_range_validity}}~\cite{Yu3,Yu2021}
\begin{equation}
\label{Yu3_reexp}
k_L(\Tilde r_j) Y_L^M(\Hat{\mathbf r}_j) = \sum\nolimits_{l_1,m_1}\mathcal H_{l_1 m_1}^{L M}(\mathbf{R}_{i j}) i_{l_1}(\Tilde r_i) Y_{l_1}^{m_1}(\Hat{\mathbf r}_i) 
\end{equation}
with re-expansion coefficients originally defined in \cite{Yu3}~as
\begin{equation}
\label{coeffs_HlmLM_definition}
\!\!\!\begin{aligned}
\mathcal H_{l_1 m_1}^{L M}(\mathbf{R}_{i j}) & \mathrel{:=} \sum_{l_2,m_2}\! (-1)^{l_1+l_2} H_{l_1 m_1 l_2 m_2}^{L M} k_{l_2}(\Tilde R_{i j}) Y_{l_2}^{m_2}(\Hat{\mathbf R}_{i j}), \\
H_{l_1 m_1 l_2 m_2}^{L M} & \mathrel{:=} C_{l_1 0 l_2 0}^{L 0} C_{l_1 m_1 l_2 m_2}^{L M} \sqrt{\tfrac{4\pi (2 l_1+1) (2 l_2+1)}{2 L+1}}, 
\end{aligned}
\end{equation}
where $r_i<R_{i j} = \|\mathbf R_{i j}\|$, $\mathbf R_{i j}=\mathbf x_j-\mathbf x_i$ points from $\mathbf x_i$ to $\mathbf x_j$, $\Tilde R_{i j} \mathrel{:=} \kappa R_{i j}$, $\Hat{\mathbf R}_{i j} \mathrel{:=} \mathbf R_{i j}/R_{i j}$, $C_{l_1 m_1 l_2 m_2}^{L M} = \left<l_1 l_2; m_1 m_2 \mid L M\right>$ are Clebsch-Gordan coefficients\footnote{\label{footnote_ClebshGordan_properties}Note that $C_{l_1 0 l_2 0}^{L 0}$ can only be nonzero if $L+l_1+l_2$ is even (thus in fact one can also put $(-1)^L$ instead of $(-1)^{l_1+l_2}$ in the original definition \eqref{coeffs_HlmLM_definition} of $\mathcal H_{l_1 m_1}^{L M}(\mathbf{R}_{i j})$), and $C_{l_1 m_1 l_2 m_2}^{L M}$ can only be nonzero when $M=m_1+m_2$, $|l_1-L|\le l_2\le l_1+L$, which clearly limits the indices $l_2$ and $m_2$ in the sum $\sum_{l_2, m_2}$ in~\eqref{coeffs_HlmLM_definition} to a finite range.}, the boundary conditions \eqref{Lin_eq_standard_bc} yield 
\begin{subequations}
\label{implement_bcs}
\begin{align}
&\begin{aligned}
&\Tilde a^n_i L_{n m,i} + \Hat L_{n m,i}/\Tilde a_i^{n+1} - k_n(\Tilde a_i) G_{n m,i} \\ 
&\  = i_n(\Tilde a_i)\sum\nolimits_{j=1,\, j\ne i}^N\sum\nolimits_{L,M}\mathcal H_{n m}^{L M}(\mathbf R_{i j}) G_{L M,j} ,
\end{aligned}\label{implement_bc1} \\
&\begin{aligned}
&\varepsilon_i n \Tilde a^{n-1}_i L_{n m,i} - (n+1)\varepsilon_i\Hat L_{n m,i}/\Tilde a_i^{n+2} - \varepsilon_{\text{sol}} k_n'(\Tilde a_i) G_{n m,i} \\ 
&\  = \varepsilon_{\text{sol}} i_n'(\Tilde a_i)\sum\nolimits_{j=1,\, j\ne i}^N\sum\nolimits_{L,M}\mathcal H_{n m}^{L M}(\mathbf R_{i j}) G_{L M,j},
\end{aligned}\label{implement_bc2}
\end{align}
\end{subequations}
$\Tilde a_i \mathrel{:=} \kappa a_i$, from which substituting \eqref{implement_bc1} into \eqref{implement_bc2} and expressing Bessel functions derivatives $k_n'(\Tilde a_i)$ and $i_n'(\Tilde a_i)$ through \eqref{diff_modifiedBessel}, we can completely exclude coefficients $\{L_{n m,i}\}$ of~\eqref{Lin_eqs_Phi_in}:
\begin{equation}
\label{main_eq_G}
\!\!\!\!\!\!
\begin{aligned}
& \bigl(n(\varepsilon_i-\varepsilon_{\text{sol}}) k_n(\Tilde a_i) + \Tilde a_i \varepsilon_{\text{sol}} k_{n+1}(\Tilde a_i)\bigr) G_{n m,i} \\
&\ + \bigl(n(\varepsilon_i-\varepsilon_{\text{sol}}) i_n(\Tilde a_i) - \Tilde a_i \varepsilon_{\text{sol}} i_{n+1}(\Tilde a_i)\bigr)\\
&\ \times \!\sum_{j=1, j\ne i}^N \sum_{L,M} \mathcal H_{n m}^{L M}(\mathbf R_{i j}) G_{L M,j} = (2 n+1)\varepsilon_i \Hat L_{n m,i}/\Tilde a_i^{n+1} .
\end{aligned}    
\end{equation}
Relations \eqref{main_eq_G} can be recast in matrix form~as
\begin{equation}
\label{eqs_G_intermediate_}
\mathsf A_i \Tilde{\mathbf G}_i + \sum\nolimits_{j=1,\, j\ne i}^N \mathsf B_{i j}\Tilde{\mathbf G}_j = \mathbf S_i ,\quad \forall i=\overline{1,\ldots,N},
\end{equation}
where we have introduced (infinite-size) column-vectors $\Tilde{\mathbf G}_i$, $\mathbf S_i$, and matrices $\mathsf A_i$, $\mathsf B_{i j}$, such that 
\begin{equation}
\label{various_matrix_definitions_0}
\begin{aligned}
\Tilde{\mathbf G}_i &\mathrel{:=} \left\{\Tilde G_{n m,i} \right\}_{n m} , \\ 
\mathbf S_i &\mathrel{:=} \left\{(2 n+1)\varepsilon_i\Hat L_{n m,i}/\Tilde a_i^{n+2} \right\}_{n m} ,\\
\mathsf A_i &\mathrel{:=} \operatorname{diagonal}\left\{\alpha_n(\Tilde a_i,\varepsilon_i) \Upsilon_{n,i}\right\}_{n m} , \\
\mathsf B_{i j} &\mathrel{:=} \left\{\beta_{n m, L M}(\Tilde a_i,\varepsilon_i,\mathbf R_{i j})\Upsilon_{L,j}\right\}_{n m, L M} 
\end{aligned}
\end{equation}
with 
\begin{subequations}
\label{alpha_beta_definitions}
\begin{align}
& \alpha_n(\Tilde a_i,\varepsilon_i) \mathrel{:=} (\varepsilon_i-\varepsilon_\text{sol}) n \frac{k_n(\Tilde a_i)}{\Tilde a_i} + \varepsilon_\text{sol} k_{n+1}(\Tilde a_i), \label{alpha_definition} \\
&\begin{aligned}
\beta_{n m, L M}(\Tilde a_i,\varepsilon_i,\mathbf R_{i j}) \mathrel{:=} & \Bigl(\!\!(\varepsilon_i\!-\!\varepsilon_\text{sol}) n \frac{i_n(\Tilde a_i)}{\Tilde a_i} \!-\! \varepsilon_\text{sol} i_{n+1}(\Tilde a_i)\!\!\Bigr) \\ 
& \times \mathcal H_{n m}^{L M}(\mathbf R_{i j}) \, ;
\end{aligned} \label{beta_definition}
\end{align}
\end{subequations}
$0\le | m |\le n$ and $0\le | M |\le L$ enumerate rows and columns in the above matrices, respectively. In \eqref{eqs_G_intermediate_} and \eqref{various_matrix_definitions_0}, we also introduced the scaling of coefficients of \eqref{Lin_eqs_Phi_out}, namely we put 
\begin{equation}
\label{G_nm_scaling}
G_{n m,i} = \Tilde G_{n m,i} \Upsilon_{n,i} \ \ \;\text{with}\ \ \;\Upsilon_{n,i} \mathrel{:=} \left((2 n+1) k_n(\Tilde a_i) a_i\right)^{-1} ;
\end{equation}
this will help us to conveniently consider $\{\Tilde G_{n m,i}\}$ as spherical Fourier coefficients of $\left.\Phi_{\text{out},i}\right|_{\partial\Omega_i}\in H^1(\partial\Omega_i)$ in the appropriate basis and to ensure the boundedness of the corresponding operators emerging in this situation and acting on the spaces of potential coefficients (see details in the proof of Proposition~\ref{proposition_matrix_repr}). Note that diagonal matrix $\mathbb A$ has always strictly positive elements on its main diagonal (see \textcolor{red}{\cite{supplem_pre}}). Matrix equations \eqref{eqs_G_intermediate_} lead to a global block system whose blocks are (infinite-size themselves) components~\eqref{various_matrix_definitions_0}:
\begin{equation}
\label{global_lin_sys1}
(\mathbb I + \mathbb K) \Vec{\Tilde{\mathbb G}} = \mathbb A^{-1} \Vec{\mathbb S}, 
\end{equation}
where block matrices $\mathbb A \mathrel{:=} \operatorname{diagonal}\{\mathsf A_i\}_{i=1}^N$, $\mathbb B \mathrel{:=}\{\mathsf B_{i j}\}_{i,j=1;\; i\ne j}^N$, $\mathbb K \mathrel{:=} \mathbb A^{-1} \mathbb B$, $\mathbb I$ is the identity matrix, and block column-vectors $\Vec{\Tilde{\mathbb G}} \mathrel{:=} \{\Tilde{\mathbf G}_i\}_{i=1}^N$, $\Vec{\mathbb S} \mathrel{:=} \{\mathbf S_i\}_{i=1}^N$; these $\mathbb K$ and $\mathbb A^{-1} \Vec{\mathbb S}$  apparently take the following expanded form:
\begin{gather*}
\mathbb K \!=\!\!
\begin{pmatrix}
\!0 \!\!&\!\! \mathsf A_1^{-1} \mathsf B_{1 2} \!\!&\!\! \ldots \!\!&\!\! \mathsf A_1^{-1} \mathsf B_{1 N} \!\!\\
\!\mathsf A_2^{-1} \mathsf B_{2 1} \!\!&\!\! 0 \!\!&\!\! \ldots \!\!&\!\! \mathsf A_2^{-1} \mathsf B_{2 N} \!\!\\
\!\vdots \!\!&\!\! \vdots \!\!&\!\! \ddots \!\!&\!\! \vdots \!\!\\
\!\mathsf A_N^{-1} \mathsf B_{N 1} \!\!&\!\! \mathsf A_N^{-1}\mathsf B_{N 2} \!\!&\!\! \ldots \!\!&\!\! 0\!\!
\end{pmatrix}\!\!, 
\ \ \   
\mathbb A^{-1} \Vec{\mathbb S} \!=\!\!
\begin{pmatrix}
\!\!\mathsf A_1^{-1} \mathbf S_1\!\! \\
\!\!\mathsf A_2^{-1} \mathbf S_2\!\! \\
\!\!\vdots\!\! \\
\!\mathsf A_N^{-1} \mathbf S_N\!\!
\end{pmatrix}\!\!
\end{gather*}
(see the joint papers \textcolor{red}{\cite{supplem_prl,supplem_pre,supplem_pre_force}} for more detailed discussions of \eqref{global_lin_sys1} and the screening-ranged expansions it infers).
\begin{remark}
\label{free_surf_charge_remark}
In the case of nonzero $\sigma_i^\text{f}\in L^2(\partial\Omega_i)$ in \eqref{Lin_eq_standard_bc_2nd}, an additional addend $\sigma_i^\text{f}/\varepsilon_0$ has to be appended to the right-hand side $\mathfrak s_i(\mathbf r)$  of \eqref{main_eq_G_op} (subsequently, to be also taken into account in $\Vec{\mathfrak s}$ of \eqref{main_eq_G_op_L2} and \eqref{main_eq_G_op_Sobolev}). Next, expanding inhomogeneous surface density $\sigma_i^\text{f}$ in Fourier series in spherical harmonics, that is $\sigma_i^\text{f}(\Hat{\mathbf r}_i) = \sum_{n,m} \sigma_{n m,i}^\text{f} Y_n^m(\Hat{\mathbf r}_i)$ with expansion coefficients 
\begin{equation}
\label{sigma_nm_i_def}
\sigma_{n m,i}^\text{f} = a_i^{-2} \oint_{\partial\Omega_i} \sigma_i^\text{f}(\Hat{\mathbf s}_i) Y_n^m(\Hat{\mathbf s}_i)^\star d s_i ,
\end{equation}
we then obtain that just an additional addend $\sigma_{n m,i}^\text{f}a_i/\varepsilon_0$ must be appended to the right-hand side of \eqref{main_eq_G}; accordingly, $\mathbf S_i$ defined in \eqref{various_matrix_definitions_0} then also switches~to 
\begin{equation}
\label{F_for_surface_charge_vec_}
\mathbf S_i = \biggl\{\frac{(2 n+1)\varepsilon_i\Hat L_{n m,i}}{\Tilde a_i^{n+2}} + \frac{\sigma_{n m,i}^\text{f}}{\varepsilon_0\kappa}\biggr\}_{n m} .
\end{equation}
\end{remark}

\subsection{Spectral analysis of the associated NP-type operators}
\label{spectral_properties_subsection}
\noindent
Let us firstly establish connections between the matrix formalism of Sec.~\ref{spherical_Fourier_coeffs_subsec} and layered potentials formulations proposed in Sec.~\ref{basic_BIE_formulations}; by doing so we will also gain another view on the proposed matrix formalism, helping us to retrieve the useful spectral properties of (infinite-size) matrix~$\mathbb K$ that are out of reach by conventional methods (e.g.~numerical calculations show that column-wise/row-wise sums of element moduli of $\mathbb K$ can take values much greater than $1$, thus standard methods of infinite matrix analysis or localization theorems like those of Gershgorin circles \cite{Cooke1950,RossSIAM,RossAustrMathBull,HornJohnson,Tretter2008,CLS_siam} are unlikely to be useful here). To this end, let us first introduce a separable Hilbert space $\pmb l^2 \mathrel{:=} \bigoplus_{i=1}^N l^2(\{ \Tilde G_{n m,i} \}_{0\le |m|\le n})$ that represents the composite space of square-summable sequences of spherical Fourier coefficients $\Tilde G_{n m,i}$ standing for the DH potentials~$\Vec{\Phi}_{\text{out}}^{\partial\Omega}$. Then, we will prove the following propositions, which are the main results of this paper and which constitute the mathematical foundation to the analytical formalism of screening-ranged expansions constructed in our study (see the joint papers \textcolor{red}{\cite{supplem_prl,supplem_pre,supplem_pre_force}}); the corresponding proofs are enclosed in a separate Sec.~\ref{Proofs_appendix_all_proposition}.
\begin{proposition}
System \eqref{global_lin_sys1} is an $\pmb l^2$-representation of operator identity~\eqref{main_eq_G_op_Sobolev} in the basis $\bigoplus_{i=1}^N\bigl\{\frac{Y_n^m(\Hat{\mathbf r}_i)}{(2 n+1)a_i} \bigr\}_{0\le|m|\le n}$ of $\mathbf H^1$; respectively, matrix $\mathbb K$ represents operator~$\mathring{\mathcal K}$.\label{proposition_matrix_repr}
\end{proposition}
\begin{proposition}
Operators  $\mathcal K \colon \mathbf L^2\to\mathbf L^2$,  $\mathring{\mathcal K} \colon \mathbf H^1\to\mathbf H^1$  and  $\mathbb K \colon \pmb l^2\to \pmb l^2$ are compact\footnote{Compactness provides a strong characterization of the operator's spectrum structure (e.g.~the spectrum of a compact operator is (at most) countable for which zero is the only possible limit point, and each nonzero point in the spectrum is an eigenvalue -- see Fact~C\ref{Compact_fact_spectrum} in Sec.~\ref{appendix_compact_fredholm_operators_summary}) and will be crucial for our further spectral analysis in the proof of Proposition~\ref{proposition_spectral_radius}.}. Operators $\mathcal K$ and $\mathring{\mathcal K}$ are topologically equivalent\footnote{For rigorous definitions of topological and isometric equivalences between operators, see Sec.~\ref{appendix_compact_fredholm_operators_summary}.}, and operators $\mathring{\mathcal K}$ and $\mathbb K$ are isometrically equivalent.\label{proposition_K_compact}
\end{proposition}

Denote the spectral radius of an operator by $r(\cdot)$.
\begin{proposition}
The following inequalities are valid for spectral radii\footnote{Proposition~\ref{proposition_spectral_radius} is rigorously proved in Sec.~\ref{Proof_appendix_spectrum_proposition} under the additional condition $\varepsilon_i \le \varepsilon_\text{sol}$ ($1\le i\le N$) representing a situation of primary interest for biomolecular sciences. However, in view of the reasoning presented below in Remark~\ref{Remark_on_proof_epsiloni_less_epsilonm} one may presume the validity of Proposition~\ref{proposition_spectral_radius} even without this condition; the corresponding rigorous considerations are more complicated and require a subtler mathematical treatment to be given elsewhere (however we provide some numerical illustrations in Sec.~\ref{section_numeric} below).\label{footnote_epsilons}}: $r(\mathring{\mathcal K})<1$, $r(\mathcal K)<1$, and $r(\mathbb K)<1$.\label{proposition_spectral_radius}
\end{proposition}
\begin{corollary}
Operator series $\sum_{\ell=0}^{+\infty}(-1)^\ell\mathcal K^\ell$, $\sum_{\ell=0}^{+\infty}(-1)^\ell\mathring{\mathcal K}^\ell$ and $\sum_{\ell=0}^{+\infty}(-1)^\ell\mathbb K^\ell$ are convergent (absolutely, in the strong (operator-space-inherent) norm).\label{corollary_Neumann_covergence}
\end{corollary}
\begin{corollary}
Solutions to systems \eqref{main_eq_G_op_L2}, \eqref{main_eq_G_op_Sobolev} and \eqref{global_lin_sys1} exist and are unique; they can be expressed by the corresponding series $\Vec{\mathfrak g} = \sum_{\ell=0}^{+\infty}(-1)^\ell\mathcal K^\ell\mathcal A^{-1}\Vec{\mathfrak s}$,  $\Vec{\Phi}_{\text{\textnormal{out}}}^{\partial\Omega} = \sum_{\ell=0}^{+\infty}(-1)^\ell\mathring{\mathcal K}^\ell\mathcal S^\kappa \mathcal A^{-1}\Vec{\mathfrak s}$ and $\Vec{\Tilde{\mathbb G}} = \sum_{\ell=0}^{+\infty}(-1)^\ell\mathbb K^\ell\mathbb A^{-1} \Vec{\mathbb S}$, converging (absolutely) in the spaces $\mathbf L^2$, $\mathbf H^1$ and $\pmb l^2$, respectively.
\label{corollary_solution_exist_unique}
\end{corollary}
\begin{corollary}
Provided that (``input") potential \eqref{varPhi_in_i_multipoles} is real, all (``output"\footnote{In the sense that their expansion coefficients are determined through equations with, in general, complex-valued coefficients, and the right-hand side that depends on the ``input" potential \eqref{varPhi_in_i_multipoles} -- see relations~\eqref{various_matrix_definitions_0}-\eqref{global_lin_sys1}.}) potentials~\eqref{Lin_eqs_Phi_in_out} are also real.\label{corollary_solutions_real}
\end{corollary}
\begin{remark}
\label{remark_operator_neumann_series}
Corollaries \ref{corollary_Neumann_covergence} and \ref{corollary_solution_exist_unique} are of the greatest practical interest and are the key result of the spectral analysis carried out here, since they mathematically enable and justify the construction of screening-ranged expansions of electrostatic quantities proposed in the joint papers~\textcolor{red}{\cite{supplem_prl,supplem_pre,supplem_pre_force}} and technically based on the operator series~$\sum_{\ell=0}^{+\infty}(-1)^\ell\mathbb K^\ell$. Indeed, denoting $\Vec{\Tilde{\mathbb G}}^{(\ell)} \mathrel{:=} (-1)^\ell\mathbb K^\ell\mathbb A^{-1} \Vec{\mathbb S}$, so that 
\begin{equation}
\label{G_neumann_series0}
\Vec{\Tilde{\mathbb G}} = \sum\nolimits_{\ell=0}^{+\infty}\Vec{\Tilde{\mathbb G}}^{(\ell)},
\end{equation}
and inheriting the block column-wise structure of vector $\Vec{\Tilde{\mathbb G}}$, that is representing $\Vec{\Tilde{\mathbb G}}^{(\ell)}$ as block column-vector $\Vec{\Tilde{\mathbb G}}^{(\ell)} = (\Tilde{\mathbf G}_i^{(\ell)})_{i=1}^N$ composed of individual (infinitely-sized) column-vectors $\Tilde{\mathbf G}_i^{(\ell)} = \{\Tilde G_{n m,i}^{(\ell)}\}_{n m}$, we derive a screening-ranged expansion of the form
\begin{equation}
\label{scr_ranged_expansion_G}
\Tilde G_{n m,i} = \sum\nolimits_{\ell=0}^{+\infty}\Tilde G_{n m,i}^{(\ell)}
\end{equation}
for the coefficients of \eqref{Lin_eqs_Phi_out}. Plugging expansion \eqref{scr_ranged_expansion_G} into \eqref{implement_bc1} we further derive the screening-ranged expansion also for the coefficients of~\eqref{Lin_eqs_Phi_in}: 
\begin{equation}
\label{scr_ranged_expansion_L}
L_{n m,i} = \sum\nolimits_{\ell=0}^{+\infty} L_{n m,i}^{(\ell)} ,
\end{equation}
which is turn is used to derive the energy expansion (see \textcolor{red}{\cite{supplem_pre}} for details)
\begin{equation}
\label{energy_expansion_components_abs_gen}
\mathcal E = \sum\nolimits_{\ell=0}^{+\infty}\mathcal E^{(\ell)} .
\end{equation}
Superscript $\ell$ in the right-hand sides of \eqref{scr_ranged_expansion_G}-\eqref{energy_expansion_components_abs_gen} indicates the order of screening (by Debye screening factors) of the addends of these series. Detailed expressions for $\Tilde G_{n m,i}^{(\ell)}$, $L_{n m,i}^{(\ell)}$ and $\mathcal E^{(\ell)}$ are derived and discussed in the joint paper~\textcolor{red}{\cite{supplem_pre}}, while in the current work we only prove the \emph{absolute convergence} of the series on the right-hand sides of \eqref{scr_ranged_expansion_G}-\eqref{energy_expansion_components_abs_gen} (which is needed for mathematically justifying algebraic manipulations with these series and specific rearrangements of their terms leveraged in~\textcolor{red}{\cite{supplem_pre, supplem_pre_force}}) -- see Sec.~\ref{appendix_scr_rang_expansions_G_L_converg}.
\end{remark}

\section{Numerical examples}
\label{section_numeric}
\begin{figure}
\subfloat[\label{30spheres_scheme}Spheres and their dielectric constants (the color of each sphere corresponds to the value of its dielectric constant according to the colormap on the right).]
{
\includegraphics[trim=4.0cm 0.09cm 3.0cm 0.75cm,clip=true,width=0.9\linewidth]{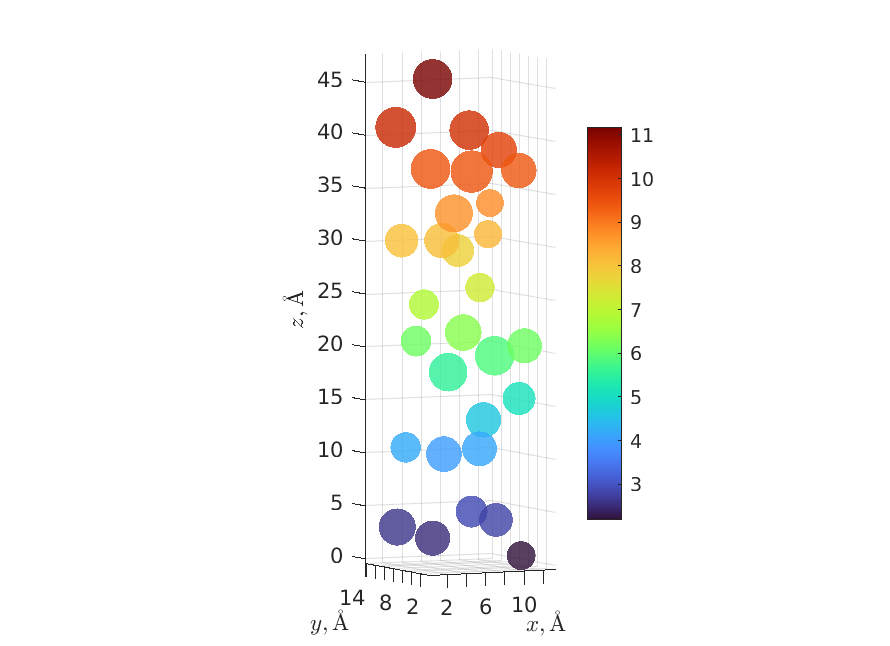}
}\hfill
\subfloat[\label{30spheres_specrad}Dependence of spectral radius $r(\mathbb K^{/n_\text{max}/})$ on $n_\text{max}$. (Lines are used to guide the eye.)]
{
\includegraphics[trim=0.6cm 0cm 1.cm 0.45cm,clip=true,width=0.95\linewidth]{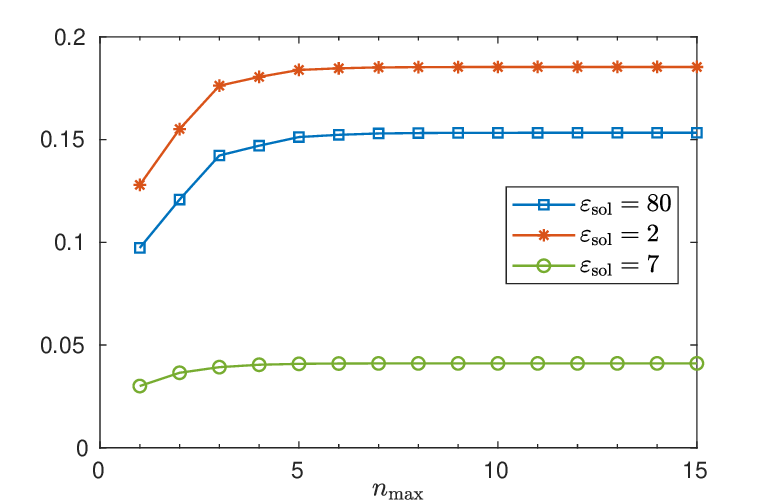}
}
\caption{A system of randomly located spheres of variable radii and dielectric constants.}
\label{30spheres_figure}
\end{figure}
\noindent
To numerically illustrate the conjecture of footnote~\ref{footnote_epsilons} (where it was stated that the spectral radii of the constructed NP-type operators is less than unity regardless of the ratios of dielectric constants $\varepsilon_i$, $\varepsilon_\text{sol}$) let us consider a system consisting of 30 randomly located non-overlapping spheres of variable radii and dielectric constants (see Fig.~\ref{30spheres_scheme}); the $\varepsilon_i$ values increase with increasing height $z$ and are determined as $\varepsilon_i=2+z_i/(5\text{\AA})$ ($z_i$ is the $z$-coordinate of the $i$-th sphere's center $\mathbf x_i$; so that $\varepsilon_i$ range between approximately 2.18 and 11.16 in this example), while for $\varepsilon_\text{sol}$ we consider the cases $\varepsilon_\text{sol}=80>\varepsilon_i$ $\forall i$, $\varepsilon_\text{sol}=2<\varepsilon_i$ $\forall i$, and finally the intermediate $\varepsilon_\text{sol}=7$ (so that there are spheres with both larger and smaller dielectric constants). By limiting the degree of spherical harmonics in \eqref{Lin_eqs_Phi_in_out} (and naturally in the remaining relations of Sec.~\ref{spherical_Fourier_coeffs_subsec}) from above by some user-defined threshold $n_\text{max}$, i.e.~by requiring $n\le n_\text{max}$, we thus obtain a finite-dimensional (finite-size) approximation $\mathbb K^{/n_\text{max}/}$ for the truly infinite-dimensional matrix $\mathbb K$. Then the spectral radius of $\mathbb K^{/n_\text{max}/}$ depending on $n_\text{max}$ is depicted in Fig.~\ref{30spheres_specrad} -- as we see, it always holds $r(\mathbb K^{/n_\text{max}/})<1$ and rapid monotonic convergence is observed with increasing~$n_\text{max}$.

\begin{figure}
\subfloat[\label{125spheres_scheme}Arrangement of spheres at $R=1$~\AA.]
{
\includegraphics[trim=2.3cm 0.25cm 2.7cm 0.7cm,clip=true,width=0.95\linewidth]{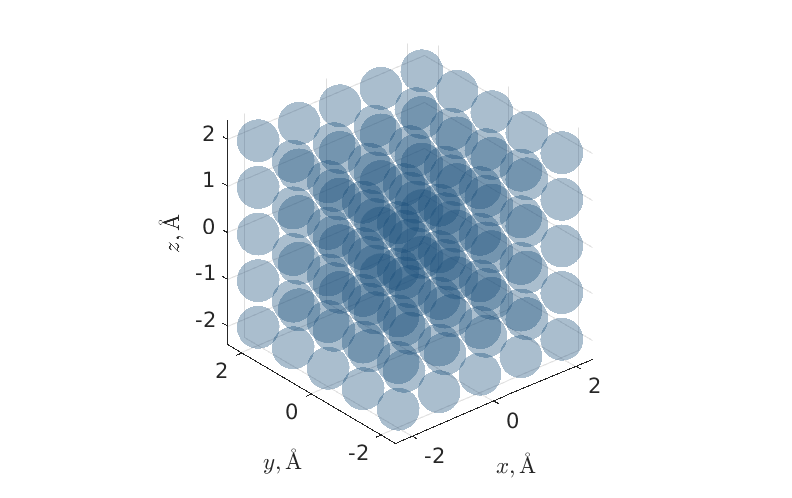} 
}\hfill
\subfloat[\label{125spheres_specrad}Dependence of spectral radius $r(\mathbb K^{/n_\text{max}/})$ on $R$. (Lines are used to guide the eye.)]
{
\includegraphics[trim=0.75cm 0.0cm 1.05cm 0.4cm,clip=true,width=0.95\linewidth]{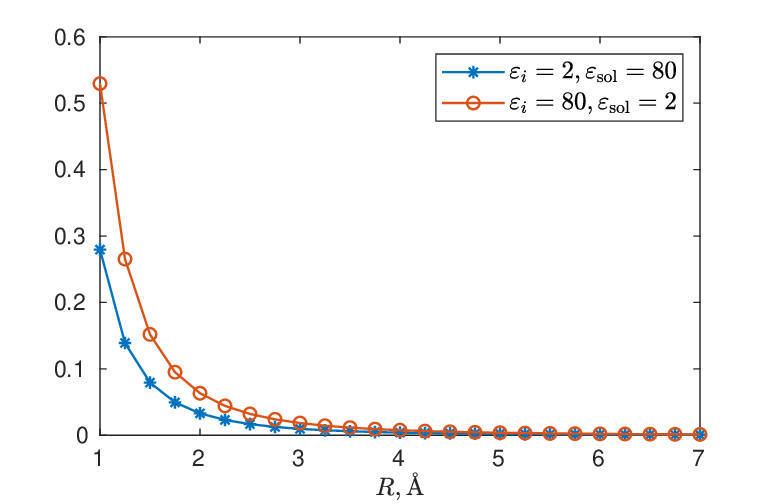} 
}
\caption{Cubic lattice of non-overlapping spheres.}
\label{125spheres_figure}
\end{figure}

In the second example, we numerically illustrate the fact that $\mathbb K$ dwindles as particles move away from each other (see Remark~\ref{Remark_on_proof_epsiloni_less_epsilonm} for details on the asymptotics of the matrix elements of $\mathbb K$). Let us consider a regular cubic lattice of 125 non-overlapping spheres of radius {0.4~\AA} centered at points $(R i', R j', R k')$, where $i', j', k' = \overline{-2, \ldots, 2}$ and $R$ (the distance between two adjacent spheres' centers in a coordinate direction) varies from {1~\AA} to {7~\AA} with a step of $0.25$~\AA; the lattice at $R=1$~\AA~is schematized in Fig.~\ref{125spheres_scheme}. The dependence of spectral radius $r(\mathbb K^{/n_\text{max}/})$ on $R$ (at fixed $n_\text{max}=10$) is then illustrated in Fig.~\ref{125spheres_specrad} -- a rapid decrease in $r(\mathbb K^{/n_\text{max}/})$ is observed as the particles move away from each other.

The default value of $\kappa$ used in the above calculations was $\kappa^{-1}= 8.0714$~\AA~(as per aqueous solutions of biophysical interest it is 0.145~M physiological NaCl concentration of monovalent species at a room temperature of 25~$^\circ$C).

\section{Proofs of the propositions}
\label{Proofs_appendix_all_proposition}
\noindent
In this section we will prove the claims formulated in Sec.~\ref{spectral_properties_subsection}.
\subsection{Mathematical preliminaries and a brief summary of facts on compact and Fredholm operators used in the proofs}
\label{appendix_compact_fredholm_operators_summary}
\noindent
In our further proofs we will extensively use some known facts from functional analysis (especially, from operator theory) \cite{HelemskyAMS,SauterSchwab}. Henceforth $B(X,Y)$ denotes the Banach space of linear \emph{bounded} (continuous) operators\footnote{See e.g.~\cite{HelemskyAMS} for definitions of \emph{bounded} and \emph{compact} operators, as well as other basic concepts adopted in functional analysis and used here. Note also that throughout the text we will always deal only with linear operators.} $X\to Y$ for Banach spaces $X$ and $Y$, equipped with the usual operator norm (i.e.~$\forall T\in B(X,Y)$ one has $\|T\| = \|T\|_{B(X,Y)} \mathrel{:=} \sup_{0\ne x\in X}\|T x\|_Y/\|x\|_X$), while $K(X,Y)$ denotes the space of \emph{compact} operators $X\to Y$; note also that $K(X,Y)$ forms a closed (in the operator norm topology) subspace of $B(X,Y)$. Also, for brevity $B(X) \mathrel{:=} B(X,X)$, $K(X) \mathrel{:=} K(X,X)$. 

Further, for any bijective\footnote{We use the terms \emph{injection}, \emph{surjection}, and \emph{bijection} in their conventional set-theoretic sense (see e.g.~\cite{HelemskyAMS}).} $T\in B(X,Y)$, Banach inverse mapping theorem guarantees that $\exists T^{-1}\in B(Y,X)$; following \cite{HelemskyAMS} we will call such operators \emph{topological isomorphisms} (or \emph{isometrical isomorphisms} if, moreover, the norms are also preserved, i.e.~$\|T x\|_Y = \|x\|_X$ $\forall x\in X$). 

Recall \cite{HelemskyAMS} that a point $\lambda\in\mathbb C$ is called a regular point of an operator $T\in B(X)$ if the operator $T-\lambda I$ (where $I\in B(X)$ is the identity operator) is a topological isomorphism, otherwise it is called a singular point of $T$; respectively, the set of singular points is called the \emph{spectrum} of $T$. The spectrum of any $T\in B(X)$ is a non-empty compact set in the complex plane~\cite{HelemskyAMS}. 

Next, two bounded operators $T_1 \colon X\to X$ and $T_2\colon Y\to Y$ are called \emph{topologically (isometrically) equivalent} if there exists a topological (isometrical) isomorphism $T \colon X\to Y$ such that equality $T T_1 = T_2 T$ holds\footnote{In other words, the diagram $$\begin{CD}
\smash{X} @>\smash{T_1}>> \smash{X} \\
@VV\smash{T}V @ VV\smash{T}V \\
\smash{Y} @>\smash{T_2}>> \smash{Y}
\end{CD}$$ is commutative (the final result of following the arrows on the diagram does not depend on the specific path one follows).}. The advantage of this concept is that equivalent operators, although possibly acting in different spaces, share out various useful properties/invariants (like spectrum or compactness) and can thus be mutually identified in this sense -- so that one can gain information about one operator by studying another (sometimes simpler)~one.

The important facts about compact operators we will use in further (especially in order to prove Propositions~\ref{proposition_K_compact} and~\ref{proposition_spectral_radius}, and Corollaries) are that~\cite{HelemskyAMS,SauterSchwab}:
\newcounter{Compact_list_counter}
\begin{list}{\bf C\arabic{Compact_list_counter}}
{\usecounter{Compact_list_counter}}
    \item\label{Compact_fact_spectrum}\emph{(Riesz-Schauder theorem)} The spectrum of an operator $T\in K(X)$ is (at most) countable for which zero is the only possible limit point, and each non-zero point in the spectrum is an eigenvalue.
    \item\label{Compact_fact_ideal}If $T_1\in B(X,Y)$ and $T_2\in B(Y,Z)$ and at least one of these operators $T_i$ is compact, then $T_2 T_1\in K(X,Z)$.
    \item\label{Compact_finite_rank}Any \emph{finite-rank} operator $T\in B(X,Y)$ (i.e.~$\operatorname{Range} T$ is finite-dimensional) belongs to $K(X,Y)$. If a sequence $\{T_n\}$ of finite-rank operators converges (in $\|\cdot\|_{B(X,Y)}$) to an operator $T\in B(X,Y)$, then $T\in K(X,Y)$; conversely, if $Y$ is a Hilbert space, then every $T\in K(X,Y)$ can be approximated (in $\|\cdot\|_{B(X,Y)}$) by a sequence of finite-rank operators.
\end{list}

Let us also recall the concept of a \emph{Fredholm} operator. An operator $A\in B(X,Y)$ is Fredholm if its kernel $\operatorname{Ker} A$ and the quotient space $Y/\operatorname{Range} A$ (i.e.~cokernel of $A$) are finite-dimensional (note that here we do not impose explicitly the frequently-used condition \cite{AmmariKang_mathstat} that subspace $\operatorname{Range} A$ is closed in $Y$, because the just imposed condition $\dim(Y/\operatorname{Range} A)<\infty$ together with \cite[Proposition~3.5.1]{HelemskyAMS} will automatically assure it); the difference $\operatorname{Ind}(A) \mathrel{:=} \dim\operatorname{Ker} A - \dim(Y/\operatorname{Range} A)$ is then called the \emph{index} of~$A$. (We will be mostly interested in operators with index zero.) Some important facts about Fredholm operators we will employ in further proofs are~\cite{HelemskyAMS}:
\newcounter{Fredholm_list_counter}
\begin{list}{\bf F\arabic{Fredholm_list_counter}}
{\usecounter{Fredholm_list_counter}}
 \item\label{Fredholm_fact_indices}If $A_1 \colon X\to Y$ and $A_2 \colon Y\to Z$ are Fredholm, $T\in K(X,Y)$, then $A_1+T \colon X\to Y$ and $A_2 A_1 \colon X\to Z$ are also Fredholm with $\operatorname{Ind}(A_1+T) = \operatorname{Ind}(A_1)$,  $\operatorname{Ind}(A_2 A_1) = \operatorname{Ind}(A_1) + \operatorname{Ind}(A_2)$.
 \item\label{Fredholm_fact_zeroindex_representation}An operator $A \colon X\to Y$ of the form $A = S+T$, where $S \colon X\to Y$ is a topological isomorphism and $T\in K(X,Y)$, is Fredholm with $\operatorname{Ind}(A)=0$. And vice versa, any zero-index Fredholm operator can be represented as such a sum (of a topological isomorphism and a compact operator).
 \item\label{Fredholm_fact_alternative}\emph{(Fredholm alternative)} A zero-index Fredholm operator is injective if and only if it is surjective.
\end{list}
Fact F\ref{Fredholm_fact_alternative} is of the most importance to us, as it provides the relatively easily verifiable condition (injectivity) to be sufficient for the stronger property of bijectivity (which, together with the Banach inverse mapping theorem \cite{HelemskyAMS, SauterSchwab}, then entails the existence of a bounded inverse); we will use it later to show the invertibility of some operators important to this study.

\begin{remark}
\label{remark_the_same_notation_potentials_surface_volumetric}
Within this section we will work with the (volumetric) single-layer potential operators $\Breve{\mathcal S}_i^\kappa \in B(H^{s-1/2}(\partial\Omega_i), H_\text{loc}^{1+s}(\mathbb R^3))$ as well as with their (boundary) counterparts (representing the appropriate surface traces, i.e.~operators of the form $\mathcal S_i^\kappa = \gamma_{\partial\Omega_i} \Breve{\mathcal S}_i^\kappa \in B(H^{s-1/2}(\partial\Omega_i), H^{s+1/2}(\partial\Omega_i))$). Here letters $H$ as usual denote Sobolev spaces (fractional-order, in general). Trace mapping $\gamma_{\partial\Omega_i} \colon  H_\text{loc}^{1+s}(\mathbb R^3)\to H^{s+1/2}(\partial\Omega_i)$ restricts its arguments to spaces on $\partial\Omega_i$; here for real $s$ one generally has $s>-1/2$ as $\partial\Omega_i\in C^\infty$ for the above-introduced trace operator $\gamma_{\partial\Omega_i}$ to be continuous, however continuous mapping properties of integral boundary operators can nevertheless be extended also for $s$ beyond this range (see comments in \cite[\S~3.1.2]{SauterSchwab}, \cite[Chap.~4]{NedelecAEE} and \cite[Chap.~7]{McLean_0}). 
While the one-sided trace operators ($\gamma_{\partial\Omega_i}^\pm$, where as before $+/-$ denotes exterior/interior) are also natural in the potential theory~\cite{SauterSchwab}, we will not use symbols like $\gamma_{\partial\Omega_i}^\pm \Breve{\mathcal S}_i^\kappa$ since in our case the potential produced by $\Breve{\mathcal S}_i^\kappa$ has a zero jump when crossing $\partial\Omega_i$ \cite{AmmariKang_polarization,VicoGreen,Chipman2004,SauterSchwab} (hence, we omitted indices $\pm$ in the definition of~$\mathcal S_i^\kappa$).

\end{remark}
\subsection{Proof of Proposition~\ref{proposition_matrix_repr}}
\label{Proof_appendix_representation_proposition}
\noindent
Let us firstly note that operators $\mathcal S_i^\kappa, \mathcal K_i^\kappa{}^\star \in K(L^2(\partial\Omega_i))$ (so Fact~C\ref{Compact_fact_spectrum} applies) and spherical harmonics \eqref{Ynm_definition} are eigenfunctions of operators $\mathcal S_i^\kappa$ and $\mathcal K_i^\kappa{}^\star$ (so that these operators are ``diagonalizable'' in this spherical basis): 
\begin{equation}
\label{eigen_functions_S_K}
\begin{aligned}
 \mathcal S_i^\kappa Y_n^m(\Hat{\mathbf r}_i) & = -\frac{\Tilde a_i^2}{\kappa} i_n(\Tilde a_i) k_n(\Tilde a_i) Y_n^m(\Hat{\mathbf r}_i),\\ 
\mathcal K_i^\kappa{}^\star Y_n^m(\Hat{\mathbf r}_i) & = -\left(\Tilde a_i^2 i_n(\Tilde a_i) k_n'(\Tilde a_i) + \frac{1}{2}\right)Y_n^m(\Hat{\mathbf r}_i)
\end{aligned}
\end{equation}
(here $\Tilde a_i$ is $\kappa a_i$ as usual; see also \eqref{diff_modifiedBessel} for the expression of $k_n'(\Tilde a_i)$). In the $\kappa\to0$ limit one has (see e.g.~asymptotics~\eqref{small_Bessel_i_k})
\begin{equation}
\label{eigen_functions_S_K_kappa_zero}
\begin{aligned}
\mathcal S_i^0 Y_n^m(\Hat{\mathbf r}_i) & = -\frac{a_i}{2 n+1} Y_n^m(\Hat{\mathbf r}_i),\\ 
\mathcal K_i^0{}^\star Y_n^m(\Hat{\mathbf r}_i) & = \frac{1}{2(2 n+1)} Y_n^m(\Hat{\mathbf r}_i).
\end{aligned}
\end{equation}
See papers \cite{VicoGreen,BardKneBru2015} for the detailed proof of \eqref{eigen_functions_S_K} (paper \cite{VicoGreen} considers similar relations in terms of the regular spherical Bessel and Hankel functions, but following the derivations in \cite{VicoGreen} and using identities \eqref{relations_modified_and_regular_Bessel_Hankel} one arrives at~\eqref{eigen_functions_S_K}). Owing to \eqref{in_kn_are_positive_} eigenvalues~\eqref{eigen_functions_S_K} of $\mathcal S_i^\kappa$ appear negative, while from \eqref{mod_bessel_asympt_large_x} one also sees that their asymptotical behavior at $n\to+\infty$ is similar to the behavior of the eigenvalues of~$\mathcal S_i^0$.

Expanding surface density $\mathfrak g_i$ (see~\eqref{vol_potentials_via_operators}) in Fourier series in spherical harmonics one immediately gets $\mathfrak g_i(\Hat{\mathbf r}_i) = \sum_{n,m} \mathfrak g_{n m,i} Y_n^m(\Hat{\mathbf r}_i)$ with the coefficients $\mathfrak g_{n m,i} = a_i^{-2} \oint_{\partial\Omega_i} \mathfrak g_i(\Hat{\mathbf s}_i) Y_n^m(\Hat{\mathbf s}_i)^\star d s_i$, which can be related to expansion coefficients \eqref{Lin_eqs_Phi_out}~as 
\begin{equation}
\label{relation_between_G_and_g}
\begin{aligned}
G_{n m,i} & = -\kappa a_i^2 i_n(\Tilde a_i) \mathfrak g_{n m,i} \\ 
& \Leftrightarrow \ \ \Tilde G_{n m,i} = -\kappa a_i^3 (2 n+1) i_n(\Tilde a_i) k_n(\Tilde a_i)\mathfrak g_{n m,i}
\end{aligned}
\end{equation}
-- indeed, it follows from the explicit expressions for $G_{n m,i}$ acquired from the expansion for the (screened) Green function for PBE in spherical coordinates~\cite{Yu3,our_jcp}: 
\begin{equation}
\label{addition_theorem_screened_0}
\frac{e^{-\kappa \|\mathbf r_i-\mathbf s_i\|}}{4\pi \|\mathbf r_i-\mathbf s_i\|} = \kappa \sum_{n,m} i_n(\Tilde a_i) k_n(\Tilde r_i) Y_n^m(\Hat{\mathbf r}_i) Y_n^m(\Hat{\mathbf s}_i)^\star,
\end{equation}
where vectors $\mathbf r_i$ and $\mathbf s_i$ are measured from the center of the $i$-th sphere according to their previous definitions, $r_i = \|\mathbf r_i\| > \|\mathbf s_i\| = a_i$; then putting expansion \eqref{addition_theorem_screened_0} into the definition \eqref{S_vol_definition} of $\Breve{\mathcal S}_i^\kappa \mathfrak g_i$ one gets
\begin{equation}
\label{Phi_i_expansion_addition_t}
\Breve{\mathcal S}_i^\kappa \mathfrak g_i (\mathbf r_i)  =  -\kappa \sum_{n,m} \left( \oint_{\partial\Omega_i} \mathfrak g_i(\Hat{\mathbf s}_i) Y_n^m(\Hat{\mathbf s}_i)^\star d s_i\right)\! i_n(\Tilde a_i) k_n(\Tilde r_i) Y_n^m(\Hat{\mathbf r}_i)
\end{equation}
(here the possibility of term-by-term integration can be easily justified using e.g.~dominated convergence theorem of Lebesgue, see~\cite[\S~2.8.2]{Olver1997} for the convenient form of it), from which the formulae for $G_{n m,i}$ (see \eqref{expansion_surface_densities_coeffs}) and their relations to $\mathfrak g_{n m,i}$ (see \eqref{relation_between_G_and_g}) immediately follow. Conversely, using $\mathfrak g_i(\Hat{\mathbf r}_i)$ given as a spherical Fourier series with the coefficients $\mathfrak g_{n m,i}$ determined in this way and then utilizing \eqref{eigen_functions_S_K} one predictably recovers the limiting value of $\Breve{\mathcal S}_i^\kappa \mathfrak g_i$ ($=\Phi_{\text{out},i}$) as $r_i\to a_i^+$, as one may expect from~\eqref{Lin_eqs_Phi_out}:
\begin{align}
& \mathcal S_i^\kappa \mathfrak g_i = \left|\text{use \eqref{relation_between_G_and_g}}\right| =  \sum_{n,m} \frac{-\kappa G_{n m,i}}{\Tilde a_i^2 i_n(\Tilde a_i)} \mathcal S_i^\kappa Y_n^m(\Hat{\mathbf r}_i) = |\text{use~\eqref{eigen_functions_S_K}}| \notag\\ 
& \!=\!\! \sum_{n,m}\!\frac{-\kappa G_{n m,i}}{\Tilde a_i^2 i_n(\Tilde a_i)}\!\frac{-\Tilde a_i^2}{\kappa}  i_n(\Tilde a_i) k_n(\Tilde a_i) Y_n^m(\Hat{\mathbf r}_i) \!=\!\! \sum_{n,m}\! G_{n m,i} k_n(\Tilde a_i) Y_n^m(\Hat{\mathbf r}_i) \notag\\  
&= \sum_{n,m} \Tilde G_{n m,i} \frac{Y_n^m(\Hat{\mathbf r}_i)}{(2 n+1) a_i} = \left.\Phi_{\text{out},i}\right|_{\partial\Omega_i}\!.\label{limiting_values_of_Phi_out_at_boundary_}
\end{align}

The systems of functions $\{Y_n^m(\Hat{\mathbf r}_i)/a_i\}_{0\le |m|\le n}$ and $\{Y_n^m(\Hat{\mathbf r}_i)/((2 n+1) a_i)\}_{0\le |m|\le n}$ will form complete orthonormal Schauder bases \cite{HelemskyAMS} of the separable Hilbert spaces $L^2(\partial\Omega_i)$ and $H^1(\partial\Omega_i)$, respectively~\cite{NedelecAEE}. Regarding the Sobolev space $H^1(\partial\Omega_i)$, at the moment we can conveniently consider it as $H^1(\partial\Omega_i)=\{u\in L^2(\partial\Omega_i),\  \nabla_{\partial\Omega_i} u\in L^2(\partial\Omega_i)^2\}$ through the surface gradient $\nabla_{\partial\Omega_i} u \mathrel{:=} \frac{1}{\sin\theta_i}\frac{\partial u}{\partial\varphi_i}\Hat{\pmb\varphi}_i + \frac{\partial u}{\partial\theta_i}\Hat{\pmb\theta}_i$ (see \cite[Eq.~(2.4.6)]{NedelecAEE}) with the inner product $(u,\, v)_{H^1(\partial\Omega_i)} = \oint_{\partial\Omega_i} u(\mathbf s) v(\mathbf s)^\star d s + 4\oint_{\partial\Omega_i} \nabla_{\partial\Omega_i}u(\mathbf s)\cdot \nabla_{\partial\Omega_i} v(\mathbf s)^\star d s $, so that using relation $\oint_{\partial\Omega_i} |\nabla_{\partial\Omega_i} Y_n^m|^2 d s = n(n+1) a_i^2$ (note that $Y_n^m$ is an eigenfunction of the Laplace-Beltrami operator on the unit sphere with the corresponding eigenvalue $-n(n+1)$, see \cite{NedelecAEE,RMCWF}) we then obtain $\|Y_n^m\|_{H^1(\partial\Omega_i)} = ((Y_n^m,\, Y_n^m)_{H^1(\partial\Omega_i)})^{1/2} = (2 n+1) a_i$. Such an inner product on $H^1(\partial\Omega_i)$ slightly differs from the conventional one, but leads to an equivalent definition of the space and an equivalent induced norm\footnote{Recall that two norms $\|\cdot\|_1$ and $\|\cdot\|_2$ on a normed space $X$ are called equivalent if $\exists c_1,c_2>0$ that $\forall x\in X$ $c_1\|x\|_1\le\|x\|_2\le c_2\|x\|_1$; equivalent norms are indistinguishable in the sense that they generate the same topology/convergence, see~\cite{HelemskyAMS}.} -- see \cite[Chap.~2]{NedelecAEE}.

Next recall that, as guaranteed by the well-known Riesz-Fischer theorem~\cite[\S~2.2]{HelemskyAMS}, $L^2(\partial\Omega_i)$ (to which $\mathfrak g_i$ belongs) is isometrically isomorphic to the space of square-summable sequences $l^2(\{ a_i \mathfrak g_{n m,i} \}_{0\le |m|\le n})$ (of Fourier coefficients of $\mathfrak g_i$ over basis $\{Y_n^m(\Hat{\mathbf r}_i)/a_i\}$). As well, the same theorem entails that $H^1(\partial\Omega_i)$ is isometrically isomorphic to the space of square-summable sequences $l^2(\{ \Tilde G_{n m,i} \}_{0\le |m|\le n})$ of Fourier coefficients $\{\Tilde G_{n m,i}\}$ of $\left.\Phi_{\text{out},i}\right|_{\partial\Omega_i}$ (which shall belong to $H^1(\partial\Omega_i)$ due to~\eqref{S_kappa_bounded_}); this actually permits us to work with matrix equations  \eqref{eqs_G_intermediate_} and \eqref{global_lin_sys1} (albeit having infinite-size matrices) to spherical Fourier coefficients instead of systems for densities~\eqref{main_eq_G_op} or surface potentials~\eqref{main_eq_G_op_Sobolev}.

Let us now analyze term-by-term the action of underlying operators in \eqref{main_eq_G_op_Sobolev}, using \eqref{eigen_functions_S_K}-\eqref{eigen_functions_S_K_kappa_zero}. First, let us note that since there holds expansion $\left.\Phi_{\text{out},j}\right|_{\partial\Omega_j} = \sum\limits_{L,M} \Tilde G_{L M,j} \frac{Y_L^M(\Hat{\mathbf r}_j)}{(2 L+1) a_j}$ (see \eqref{limiting_values_of_Phi_out_at_boundary_}), we have $$(\mathcal S_j^\kappa)^{-1} \left.\Phi_{\text{out},j}\right|_{\partial\Omega_j} = \sum\limits_{L,M} \frac{-\kappa^2 \Tilde G_{L M,j}}{(2 L+1) \Tilde a_j^3 i_L(\Tilde a_j) k_L(\Tilde a_j)} Y_L^M(\Hat{\mathbf r}_j)$$ (see \eqref{eigen_functions_S_K}), so using \eqref{Phi_i_expansion_addition_t} with the last expression in place of $\mathfrak g_i$ and the orthogonality of spherical harmonics one then readily recovers the expression stated in~\eqref{Lin_eqs_Phi_out},  $\Breve{\mathcal S}_j^\kappa (\mathcal S_j^\kappa)^{-1} \left.\Phi_{\text{out},j}\right|_{\partial\Omega_j} = -\kappa \sum\limits_{L,M}\Tilde G_{L M,j} \frac{-\kappa^2}{(2 L+1) \Tilde a_j^3 i_L(\Tilde a_j) k_L(\Tilde a_j)} a_j^2 i_L(\Tilde a_j) k_L(\Tilde r_j) Y_L^M(\Hat{\mathbf r}_j) = \sum\limits_{L,M} \Tilde G_{L M,j} \frac{k_L(\Tilde r_j)}{(2 L+1) k_L(\Tilde a_j) a_j} Y_L^M(\Hat{\mathbf r}_j) = \sum\limits_{L,M} G_{L M,j} k_L(\Tilde r_j) Y_L^M(\Hat{\mathbf r}_j)$. These equivalent algebraic manipulations are nevertheless useful for understanding the action of individual addends of operator $\mathcal B_{i j}$ when it is transformed to use boundary potentials (see \eqref{main_eq_G_op_Sobolev}) --- namely, we obtain (note that one needs to calculate it on the surface $\partial\Omega_i$  when applying operators like $\left( -\frac{1}{2}\mathcal I_i + \mathcal K_i^0{}^\star \right)(\mathcal S_i^0 )^{-1}$ -- see formulation~\eqref{main_eq_G_op}):
\begin{widetext}
\begin{flalign*}
& \varepsilon_i\left(\! -\frac{1}{2}\mathcal I_i + \mathcal K_i^0{}^\star \!\right)(\mathcal S_i^0 )^{-1} \gamma_{\partial\Omega_i} \Breve{\mathcal S}_j^\kappa (\mathcal S_j^\kappa)^{-1}\!\left.\Phi_{\text{out},j}\right|_{\partial\Omega_j} = \varepsilon_i\left(\! -\frac{1}{2}\mathcal I_i + \mathcal K_i^0{}^\star \!\right)(\mathcal S_i^0 )^{-1} \! \sum_{L,M} \frac{\Tilde G_{L M,j}}{(2 L+1) k_L(\Tilde a_j) a_j}\! \left.\left(k_L(\Tilde r_j) Y_L^M(\Hat{\mathbf r}_j)\right)\right|_{\partial\Omega_i} & &\\ 
&\quad = \left|\text{use~\eqref{Yu3_reexp}}\right| = \varepsilon_i\left(-\frac{1}{2}\mathcal I_i + \mathcal K_i^0{}^\star \right)(\mathcal S_i^0 )^{-1} \sum_{L,M} \frac{\Tilde G_{L M,j}}{(2 L+1) k_L(\Tilde a_j) a_j} \sum_{n,m} i_n(\Tilde a_i) \mathcal H_{n m}^{L M}(\mathbf{R}_{i j}) Y_n^m(\Hat{\mathbf r}_i) = \left|\text{use \eqref{eigen_functions_S_K_kappa_zero}}\right| & &\\
&\quad = -\frac{\varepsilon_i}{a_i} \sum_{L,M} \frac{\Tilde G_{L M,j}}{(2 L+1) k_L(\Tilde a_j) a_j} \sum_{n,m} i_n(\Tilde a_i) \mathcal H_{n m}^{L M}(\mathbf{R}_{i j}) (2 n+1)\frac{1}{2}\left(\frac{1}{2 n+1}-1\right) Y_n^m(\Hat{\mathbf r}_i) & &\\
& \quad =\frac{\varepsilon_i}{a_i} \sum_{L,M} \frac{\Tilde G_{L M,j}}{(2 L+1) k_L(\Tilde a_j) a_j} \sum_{n,m} i_n(\Tilde a_i) \mathcal H_{n m}^{L M}(\mathbf{R}_{i j}) n Y_n^m(\Hat{\mathbf r}_i) & & \\
\intertext{and}
& -\varepsilon_\text{sol}\frac{\partial}{\partial\mathbf n_i^+}\Breve{\mathcal S}_j^\kappa (\mathcal S_j^\kappa)^{-1} \left.\Phi_{\text{out},j}\right|_{\partial\Omega_j} = \left|\text{use~\eqref{Yu3_reexp}}\right| = -\varepsilon_\text{sol}\kappa \sum_{L,M} \frac{\Tilde G_{L M,j}}{(2 L+1) k_L(\Tilde a_j) a_j} \sum_{n,m} i_n'(\Tilde a_i) \mathcal H_{n m}^{L M}(\mathbf{R}_{i j}) Y_n^m(\Hat{\mathbf r}_i) & & \\
&\quad = -\varepsilon_\text{sol}\kappa \sum_{L,M} \frac{\Tilde G_{L M,j}}{(2 L+1) k_L(\Tilde a_j) a_j} \sum_{n,m} \left(\frac{n}{\Tilde a_i}i_n(\Tilde a_i) + i_{n+1}(\Tilde a_i)\right) \mathcal H_{n m}^{L M}(\mathbf{R}_{i j}) Y_n^m(\Hat{\mathbf r}_i). & &
\end{flalign*}
These relations show that
\begin{flalign*}
& \mathcal B_{i j} (\mathcal S_j^\kappa)^{-1} \left.\Phi_{\text{out},j}\right|_{\partial\Omega_j} = \kappa 
\sum_{L,M}\frac{\Tilde G_{L M,j}}{(2 L+1) k_L(\Tilde a_j) a_j} \sum_{n,m}\left( \frac{(\varepsilon_i-\varepsilon_\text{sol}) n}{\Tilde a_i} i_n(\Tilde a_i) - \varepsilon_\text{sol} i_{n+1}(\Tilde a_i) \right)\! \mathcal H_{n m}^{L M}(\mathbf{R}_{i j}) Y_n^m(\Hat{\mathbf r}_i) & & \\ 
&\quad = \kappa \sum_{L,M}\frac{\Tilde G_{L M,j}}{(2 L+1) k_L(\Tilde a_j) a_j} \sum_{n,m} \beta_{n m, L M}(\Tilde a_i,\varepsilon_i,\mathbf R_{i j}) Y_n^m(\Hat{\mathbf r}_i) . & &
\end{flalign*}
As well, for the components of operator $\mathcal A_i$ we obtain by virtue of \eqref{eigen_functions_S_K},~\eqref{eigen_functions_S_K_kappa_zero}:
\begin{flalign*}
& \varepsilon_i\left(-\frac{1}{2}\mathcal I_i + \mathcal K_i^0{}^\star \right)(\mathcal S_i^0 )^{-1} \mathcal S_i^\kappa Y_n^m(\Hat{\mathbf r}_i) = \varepsilon_i\left(-\frac{1}{2}\mathcal I_i + \mathcal K_i^0{}^\star \right) \frac{2 n+1}{-a_i} \frac{-\Tilde a_i^2}{\kappa} i_n(\Tilde a_i) k_n(\Tilde a_i) Y_n^m(\Hat{\mathbf r}_i) & & \\
&\quad  = \frac{\varepsilon_i}{2}\left(\frac{1}{2 n+1}-1\right)(2 n+1)\Tilde a_i i_n(\Tilde a_i) k_n(\Tilde a_i)Y_n^m(\Hat{\mathbf r}_i) = -\varepsilon_i n \Tilde a_i i_n(\Tilde a_i) k_n(\Tilde a_i) Y_n^m(\Hat{\mathbf r}_i)  & &\\
\intertext{and}
& -\varepsilon_\text{sol}\left(\frac{1}{2}\mathcal I_i + \mathcal K_i^\kappa{}^\star \right) Y_n^m(\Hat{\mathbf r}_i) = -\varepsilon_\text{sol} \left(-\Tilde a_i^2 i_n(\Tilde a_i) k_n'(\Tilde a_i)\right) Y_n^m(\Hat{\mathbf r}_i) = \varepsilon_\text{sol} \Tilde a_i^2 i_n(\Tilde a_i) \left(\frac{n}{\Tilde a_i}k_n(\Tilde a_i) - k_{n+1}(\Tilde a_i)\right) Y_n^m(\Hat{\mathbf r}_i), & &
\end{flalign*}
so that
\begin{equation}
\label{mathcal_Ai_part_1}
\mathcal A_i Y_n^m(\Hat{\mathbf r}_i) = - \Tilde a_i^2 \frac{\varepsilon_i n}{\Tilde a_i} i_n(\Tilde a_i) k_n(\Tilde a_i) Y_n^m(\Hat{\mathbf r}_i) + \varepsilon_\text{sol} \Tilde a_i^2 i_n(\Tilde a_i) \left(\frac{n}{\Tilde a_i}k_n(\Tilde a_i) - k_{n+1}(\Tilde a_i)\right) Y_n^m(\Hat{\mathbf r}_i) = - \Tilde a_i^2 i_n(\Tilde a_i) \alpha_n(\Tilde a_i,\varepsilon_i) Y_n^m(\Hat{\mathbf r}_i).
\end{equation}
Note that for all indices $0\le|m|\le n$ one has $i_n(\Tilde a_i)>0$ and $\alpha_n(\Tilde a_i,\varepsilon_i)>0$ in the right-hand side of~\eqref{mathcal_Ai_part_1}. Now, from these calculations we finally obtain for the components $\mathring{\mathcal K}_{i j} = \mathcal S_i^\kappa \mathcal K_{i j} (\mathcal S_j^\kappa)^{-1} = \mathcal S_i^\kappa \mathcal A_i^{-1}\mathcal B_{i j} (\mathcal S_j^\kappa)^{-1}$ of operator $\mathring{\mathcal K} = \{\mathring{\mathcal K}_{i j}\}_{i,j=1}^N$:
\begin{flalign}
\mathring{\mathcal K}_{i j} \left.\Phi_{\text{out},j}\right|_{\partial\Omega_j} & = \kappa 
\sum_{L,M} \frac{\Tilde G_{L M,j}}{(2 L+1) k_L(\Tilde a_j) a_j} \sum_{n,m} \beta_{n m, L M}(\Tilde a_i,\varepsilon_i,\mathbf R_{i j}) \frac{\Tilde a_i^2}{-\kappa}i_n(\Tilde a_i) k_n(\Tilde a_i)\frac{-1}{\Tilde a_i^2 i_n(\Tilde a_i) \alpha_n(\Tilde a_i,\varepsilon_i)} Y_n^m(\Hat{\mathbf r}_i) & & \notag\\
& = \sum_{L,M} \Tilde G_{L M,j} \sum_{n,m} \frac{Y_n^m(\Hat{\mathbf r}_i)}{(2 n+1) a_i} \underbrace{\left(\alpha_n(\Tilde a_i,\varepsilon_i) \Upsilon_{n,i}\right)^{-1} \beta_{n m, L M}(\Tilde a_i,\varepsilon_i,\mathbf R_{i j})\Upsilon_{L,j}}_\text{Elements of $\mathsf A_i^{-1}\mathsf B_{i j}$-block of $\mathbb K$ in~\eqref{global_lin_sys1}} .\label{mathcal_K_Sobolev_representation}
\end{flalign}

Finally, for the right-hand side in \eqref{main_eq_G_op_Sobolev} we obtain:
\begin{flalign}
&\mathfrak s_i = \left|\text{use \eqref{varPhi_in_i_multipoles}, \eqref{eigen_functions_S_K_kappa_zero}}\right|= \varepsilon_i\sum_{n,m}\frac{1}{\Tilde a_i^{n+1}}\Hat L_{n m,i}\frac{1}{2}\left(\frac{1}{2 n+1}-1\right)\frac{2 n+1}{-a_i} Y_n^m(\Hat{\mathbf r}_i) - \varepsilon_i\kappa\sum_{n,m}\frac{-(n+1)}{\Tilde a_i^{n+2}}\Hat L_{n m,i} Y_n^m(\Hat{\mathbf r}_i) = \varepsilon_i\kappa\sum_{n,m}\frac{2 n+1}{\Tilde a_i^{n+2}} \Hat L_{n m,i} Y_n^m(\Hat{\mathbf r}_i), && \notag \\
\intertext{then}
&\begin{aligned}
\mathcal S_i^\kappa \mathcal A_i^{-1}\mathfrak s_i & = \left|\text{use \eqref{eigen_functions_S_K}, \eqref{mathcal_Ai_part_1}}\right| = \varepsilon_i \kappa \sum_{n,m}\frac{\Tilde a_i^2}{-\kappa} i_n(\Tilde a_i) k_n(\Tilde a_i) \frac{-1}{\Tilde a_i^2 i_n(\Tilde a_i) \alpha_n(\Tilde a_i,\varepsilon_i)} \frac{(2 n+1)^2}{\Tilde a_i^{n+2}}\Hat L_{n m,i} a_i \frac{Y_n^m(\Hat{\mathbf r}_i)}{(2 n+1)a_i} \\
& = \sum_{n,m} \underbrace{\left(\alpha_n(\Tilde a_i,\varepsilon_i) \Upsilon_{n,i}\right)^{-1}\frac{(2 n+1)\varepsilon_i}{\Tilde a_i^{n+2}}\Hat L_{n m,i}}_\text{Elements of $\mathsf A_i^{-1} \mathbf S_i$-block of $\mathbb A^{-1}\Vec{\mathbb S}$ in~\eqref{global_lin_sys1}}\frac{Y_n^m(\Hat{\mathbf r}_i)}{(2 n+1)a_i}. 
\end{aligned}& & \label{f_Sobolev_representation}
\end{flalign}
\end{widetext}

Thus, \eqref{limiting_values_of_Phi_out_at_boundary_}, \eqref{mathcal_K_Sobolev_representation} and \eqref{f_Sobolev_representation} confirm that matrix equation~\eqref{global_lin_sys1} is a representation of identity \eqref{main_eq_G_op_Sobolev} (in particular, operators $\mathring{\mathcal K}$ and $\mathcal I$) in the basis $$\bigoplus_{i=1}^N\left\{\frac{Y_n^m(\Hat{\mathbf r}_i)}{(2 n+1)a_i} \right\}_{0\le|m|\le n}$$ of~$\mathbf H^1$. Hence, we can interpret \eqref{global_lin_sys1} as an operator equation with $\pmb l^2\to\pmb l^2$ operators $\mathbb I$ and $\mathbb K$ (specified in the canonical orthonormal basis of $\pmb l^2$ by matrices of the same names), where separable Hilbert space $\pmb l^2 \mathrel{:=} \bigoplus_{i=1}^N l^2(\{ \Tilde G_{n m,i} \}_{0\le |m|\le n})$ represents the composite space of square-summable sequences of Fourier coefficients $\Tilde G_{n m,i}$ furnishing the DH potentials~$\Vec{\Phi}_{\text{out}}^{\partial\Omega}$ (and, respectively, determining the expansion coefficients $G_{n m,i} = \Tilde G_{n m,i}\Upsilon_{n,i}=\frac{\Tilde G_{n m,i}}{(2 n+1) k_n(\Tilde a_i) a_i}$ for all potentials \eqref{Lin_eqs_Phi_out} in the whole solvent region~$\Omega_\text{sol}$).~\qed

\subsection{Proof of Proposition~\ref{proposition_K_compact}}
\label{Proof_appendix_compactness_proposition}
\noindent
Let us first prove auxiliary lemma that operator $\mathcal A_{i} \colon L^2(\partial\Omega_i)\to L^2(\partial\Omega_i)$ has a bounded inverse operator. Various facts established in the course of its proof will also be useful in the proof of the following Lemma~\ref{Bij_is_compact}.
\begin{lemma}
\label{Ai_has_bounded_inverse}
$\mathcal A_i\in B(L^2(\partial\Omega_i))$  and  $\exists\mathcal A_i^{-1}\in B(L^2(\partial\Omega_i))$.
\end{lemma}
\emph{Proof.} \textsl{Step 1. Decomposition of operator $\mathcal S_i^\kappa$}. Expanding exponential kernel of integral operator $\mathcal S_i^\kappa$ in Taylor series around zero we represent $\mathcal S_i^\kappa$~as 
\begin{equation}
\label{S_kappa_decomposition0}
\mathcal S_i^\kappa = \mathcal S_i^0 + \mathcal R_i^\kappa
\end{equation}
with a ``smoother'' integral operator $\mathcal R_i^\kappa$ having non-singular integral kernel (in fact, it is equal to $\frac{e^{-\kappa \|\mathbf r-\mathbf s\|}}{4\pi \|\mathbf r-\mathbf s\|} - \frac{1}{4\pi \|\mathbf r-\mathbf s\|} = \frac{1}{4\pi}\sum_{n=1}^{+\infty}\frac{(-1)^n\kappa^n\|\mathbf r-\mathbf s\|^{n-1}}{n!}$, which converges absolutely for arbitrary (finite) arguments). Let us note that similar decompositions are well-known for the regular Helmholtz equations (i.e.~those with the reverted sign in the front of $\kappa^2$ in \eqref{Lin_eqs_lpb} and with the solutions satisfying Sommerfeld radiation conditions) for problems in acoustics and electromagnetism, see e.g.~\cite{NedelecAEE,AmmariKang_mathstat,SauterSchwab, McCamyStephan} and references therein, but we are not aware of such decompositions for the PBE (or modified Helmholtz equation) in the form we need for our further purposes; it motivates us to describe this preliminary step in more details. Since all our boundaries $\partial\Omega_i$ are smooth (they are spheres, actually), it then follows from \cite[Theorem~7.17]{McLean_0}) that $\mathcal S_i^0 \colon  H^s(\partial\Omega_i) \to H^{s+1}(\partial\Omega_i)$ ($s\ge0$) is a zero-index Fredholm operator and $\operatorname{Ker}\mathcal S_i^0$ does not depend on $s$, on the other hand $\operatorname{Ker}\mathcal S_i^0 = \{0\}$ (see \cite[Lemma~2.25]{AmmariKang_polarization}) which indicates the injectivity of $\mathcal S_i^0$, thus Fact~F\ref{Fredholm_fact_alternative} (see the beginning of the current section) immediately yields that operator 
\begin{equation}
\label{S_0_bounded_}
\mathcal S_i^0 \in B( H^s(\partial\Omega_i) , H^{s+1}(\partial\Omega_i))
\end{equation}
is boundedly invertible, i.e.  
\begin{equation}
\label{S0_i_decomposition_A}
\exists(\mathcal S_i^0)^{-1}\in B(H^{s+1}(\partial\Omega_i),H^{s}(\partial\Omega_i)).
\end{equation}
Using advanced tools of pseudodifferential operators it is possible to extend the last result also for $s<0$ (see \cite[Sect.~III]{ChenSun_jmaa}), which may be useful, for instance, for rigorous treating of free point charges situating directly on boundaries $\partial\Omega_i$ (see e.g.~recent paper \cite{Hassan_Stamm_jctc} which considers dielectric spheres with surface free point charges and proposes effective numerical methods for such systems), but to make the proofs easier we do not require such kind of generalization here. Coming to integral operator $\mathcal R_i^\kappa$, one can show~that 
\begin{equation}
\label{R_i_decomposition_A}
\mathcal R_i^\kappa \in B(H^{s}(\partial\Omega_i),H^{s+3}(\partial\Omega_i)).
\end{equation}
Indeed, further representing its integral kernel as $\frac{-\kappa}{4\pi} + \frac{1}{4\pi}\sum_{n=2}^{+\infty}\frac{(-1)^n\kappa^n\|\mathbf r-\mathbf s\|^{n-1}}{n!}$ and denoting the corresponding associated integral operators as $^1\mathcal R_i^\kappa$ and $^2\mathcal R_i^\kappa$, one can employ the argumenting of \cite[Sect.~4]{McCamyStephan} to show that $^2\mathcal R_i^\kappa$ (whose integral kernel starts with $O(\|\mathbf r-\mathbf s\|)$) is a pseudodifferential operator of order $-3$, hereby relying on \cite[Lemma~4.1]{McCamyStephan} one gets $^2\mathcal R_i^\kappa\in B(H^{s}(\partial\Omega_i),H^{s+3}(\partial\Omega_i))$. As well, owing to the constant integral kernel $\frac{-\kappa}{4\pi}$, one immediately has $^1\mathcal R_i^\kappa\in B(H^{s}(\partial\Omega_i),H^{s+3}(\partial\Omega_i))$, and so for the sum $\mathcal R_i^\kappa ={}^1\mathcal R_i^\kappa + {}^2\mathcal R_i^\kappa$.

\textsl{Step 2. Fredholmness of $\mathcal A_{i} \colon L^2(\partial\Omega_i)\to L^2(\partial\Omega_i)$}. Employing \eqref{S_kappa_decomposition0} we receive
\begin{equation}
\label{A_kappa_decomposition}
\mathcal A_{i} = -\frac{\varepsilon_i+\varepsilon_\text{sol}}{2}\mathcal I_i + \mathcal T_i,
\end{equation}
where operator $\mathcal T_i$ is $\mathcal T_i \mathrel{:=} \varepsilon_i \mathcal K_i^0{}^\star - \varepsilon_\text{sol} \mathcal K_i^\kappa{}^\star + \varepsilon_i\left(-\frac{1}{2}\mathcal I_i + \mathcal K_i^0{}^\star \right)(\mathcal S_i^0 )^{-1} \mathcal R_i^\kappa$. The first addend of \eqref{A_kappa_decomposition} is obviously a topological isomorphism on $L^2(\partial\Omega_i)$, as being the identity operator $\mathcal I_i$ scaled by non-zero constant $-\frac{\varepsilon_i+\varepsilon_\text{sol}}{2}$. Thus, in order to use Fact~F\ref{Fredholm_fact_zeroindex_representation} we need to prove that $\mathcal T_i$ is compact in $L^2(\partial\Omega_i)$. The compactness of NP operators $\mathcal K_i^0{}^\star$ and $\mathcal K_i^\kappa{}^\star$ in $L^2(\partial\Omega_i)$ is a well-known fact~\cite{AmmariKang_mathstat}, so only the last addend of $\mathcal T_i$ remains to be considered. From \eqref{S0_i_decomposition_A} and \eqref{R_i_decomposition_A} we obtain that $(\mathcal S_i^0)^{-1}\mathcal R_i^\kappa$ maps continuously $L^2(\partial\Omega_i)$ into $H^2(\partial\Omega_i)$, which is compactly embedded into $L^2(\partial\Omega_i)$ (due to the Rellich-Kondrachov compact embedding theorem -- see \cite[\S~2.5]{SauterSchwab}, \cite[Chap.~3]{McLean_0}) so that the injective identity operator acting as $H^2(\partial\Omega_i)\to L^2(\partial\Omega_i)$ is compact. Then, using this fact, Fact~C\ref{Compact_fact_ideal} and taking into account~that 
\begin{equation}
\label{Identity+plus_K0}
\Bigl(-\frac{1}{2}\mathcal I_i + \mathcal K_i^0{}^\star \Bigr)\in B(H^s(\partial\Omega_i))
\end{equation}
as $s\ge0$ (by referring to \cite[Theorem~3.1.16]{SauterSchwab}), we eventually conclude that $\mathcal T_i\in K(L^2(\partial\Omega_i))$. Apparently, this and \eqref{A_kappa_decomposition} then immediately entail that $\mathcal A_i\in B(L^2(\partial\Omega_i))$. Then, due to decomposition \eqref{A_kappa_decomposition}, Fact~F\ref{Fredholm_fact_zeroindex_representation} assures that $\mathcal A_{i} \colon L^2(\partial\Omega_i)\to L^2(\partial\Omega_i)$ is a zero-index Fredholm operator.

\textsl{Step 3. Injectivity of $\mathcal A_{i}$}. Let us show that $\mathcal A_{i}$ is injective, i.e.~$\forall\mathfrak g_i\in L^2(\partial\Omega_i)$: $\mathcal A_{i}\mathfrak g_i = 0$ $\Rightarrow$ $\mathfrak g_i = 0$. To this end, expanding $\mathfrak g_i$ in Fourier series $\mathfrak g_i = \sum_{n,m} \mathfrak g_{n m,i} Y_n^m(\Hat{\mathbf r}_i)$ and applying \eqref{mathcal_Ai_part_1} we obtain for the inner product $$0 = (\mathcal A_{i}\mathfrak g_i, \mathfrak g_i)_{L^2(\partial\Omega_i)} = -\kappa^2 a_i^4 \sum_{n,m} i_n(\Tilde a_i) \alpha_n(\Tilde a_i,\varepsilon_i)|\mathfrak g_{n m,i}|^2 ,$$
which can be fulfilled only when all $\mathfrak g_{n m,i} = 0$ (note that for all indices $0\le|m|\le n$ one has $i_n(\Tilde a_i)>0$ and $\alpha_n(\Tilde a_i,\varepsilon_i)>0$). This shows that $\mathcal A_{i}\mathfrak g_i = 0$ entails $\mathfrak g_i = 0$.

\textsl{Step 4. Conclusion of the proof}. So far, we have shown that $\mathcal A_{i}$ is an injective zero-index Fredholm operator. Then, by virtue of Fact~F\ref{Fredholm_fact_alternative} we obtain that $\mathcal A_{i}\in B(L^2(\partial\Omega_i))$ is a bijective bounded operator, thus (by Banach inverse mapping theorem) it is boundedly invertible on~$L^2(\partial\Omega_i)$.~\qed

\begin{remark}
\label{remark_single_layer_potential_mapping_property}
As a by-product of \eqref{S_kappa_decomposition0}, \eqref{S_0_bounded_} and \eqref{R_i_decomposition_A}  one can easily obtain the following useful observation (we relied on it to derive formulation~\eqref{main_eq_G_op_Sobolev}):
\begin{equation}
\label{S_kappa_bounded_}
\begin{aligned}
& \mathcal S_i^\kappa \in B(L^2(\partial\Omega_i), H^1(\partial\Omega_i)) \qquad\text{and} \\
& \exists (\mathcal S_i^\kappa)^{-1}\in B(H^1(\partial\Omega_i), L^2(\partial\Omega_i)).
\end{aligned}
\end{equation}
The first inclusion in \eqref{S_kappa_bounded_} is widely known (see e.g.~\cite[Theorem~3.1.16]{SauterSchwab}). As for the latter, since $H^3(\partial\Omega_i)$ is compactly embedded into $H^1(\partial\Omega_i)$ (by Rellich-Kondrachov compact embedding theorem), we obtain from \eqref{R_i_decomposition_A} that $\mathcal R_i^\kappa \in K(L^2(\partial\Omega_i), H^1(\partial\Omega_i))$. Then $\mathcal S_i^\kappa \in B(L^2(\partial\Omega_i), H^1(\partial\Omega_i))$ is still a zero-index Fredholm operator (since adding a compact operator does not change the index, see Fact~F\ref{Fredholm_fact_indices}). Injectivity of $\mathcal S_i^\kappa$ can also be easily checked (e.g.~using~\eqref{eigen_functions_S_K}). Thus, by Fact~F\ref{Fredholm_fact_alternative} we arrive at~\eqref{S_kappa_bounded_}.
\end{remark}

Let us now prove that $\mathcal B_{i j} \colon L^2(\partial\Omega_j)\to L^2(\partial\Omega_i)$ is compact.
\begin{lemma}
\label{Bij_is_compact}
$\mathcal B_{i j}\in K(L^2(\partial\Omega_j), L^2(\partial\Omega_i))$.
\end{lemma}
\emph{Proof.} \textsl{Step 1.} Coming to the first operator addend in $\mathcal B_{i j}$, due to formulation \eqref{main_eq_G_op}, the result of applying $\Breve{\mathcal S}_j^\kappa$ to a density $\mathfrak g_j\in L^2(\partial\Omega_j)$ shall be calculated on surface $\partial\Omega_i$, i.e.~technically speaking one actually has the trace operator $\gamma_{\partial\Omega_i}$ (see Remark~\ref{remark_the_same_notation_potentials_surface_volumetric}) being applied to~$\Breve{\mathcal S}_j^\kappa \mathfrak g_j$. Since $\Breve{\mathcal S}_j^\kappa \mathfrak g_j$ decreases with distance from $\Omega_j$ and furthermore is infinitely differentiable inside $\mathbb R^3\setminus\partial\Omega_j$ (see \cite[Theorem~3.1.1]{SauterSchwab}), one thus can obtain the continuous mapping $\gamma_{\partial\Omega_i}\Breve{\mathcal S}_j^\kappa \mathfrak g_j \colon L^2(\partial\Omega_j)\to H^s(\partial\Omega_i)$ with an $s>1$. Now, using \eqref{S0_i_decomposition_A}, \eqref{Identity+plus_K0}, Rellich-Kondrachov compact embedding theorem, and Fact~C\ref{Compact_fact_ideal}, we conclude that operator $\varepsilon_i\left(-\frac{1}{2}\mathcal I_i + \mathcal K_i^0{}^\star \right)(\mathcal S_i^0 )^{-1} \gamma_{\partial\Omega_i} \Breve{\mathcal S}_j^\kappa \in K(L^2(\partial\Omega_j), L^2(\partial\Omega_i))$ (see the proof of Fredholmness of $\mathcal A_i$ in Lemma~\ref{Ai_has_bounded_inverse}, where similar reasoning was used, but now there will be operator $\gamma_{\partial\Omega_i}\Breve{\mathcal S}_j^\kappa$ instead of~$\mathcal R_i^\kappa$).

\textsl{Step 2.} Finally, belonging of the second operator addend in $\mathcal B_{i j}$ (i.e.~that with $\frac{\partial}{\partial\mathbf n_i^+}\Breve{\mathcal S}_j^\kappa$) to $K(L^2(\partial\Omega_j), L^2(\partial\Omega_i))$ can be easily proven by observing the fact that this addend has a smooth integral kernel. For instance, for every $\mathbf s\in\partial\Omega_j$ the integral kernel $\frac{e^{-\kappa \|\mathbf r-\mathbf s\|}}{4\pi \|\mathbf r-\mathbf s\|}$ of $\Breve{\mathcal S}_j^\kappa$ is bounded and differentiable at any point $\mathbf r\in\partial\Omega_i$ (as well as in all points $\mathbf r$ from any sufficiently small neighborhoods in $\mathbb R^3$ of that point $\mathbf r$ which always exist because balls $\Omega_i$ are non-overlapping), and for all such $\mathbf r$ the integral kernel is integrable (as well as its square) over $\partial\Omega_j$. Lebesgue theorem of dominated convergence \cite{SauterSchwab} then implies that differentiation (with respect to $\mathbf r$) and integration over $\partial\Omega_j$ can be interchanged, and such a new differentiated kernel is integrable together with its square. Hence, we obtain a compact operator (due to the smoothness of the integral kernel one can show following \cite[\S~3.4]{Ammari_Ciraolo_1} that it should also represent a Hilbert-Schmidt operator, however we will not use this fact in our further analysis).~\qed
 
\emph{Continuation of the proof of Proposition~\ref{proposition_K_compact}.} From Lemma~\ref{Ai_has_bounded_inverse}, Lemma~\ref{Bij_is_compact} and Fact~C\ref{Compact_fact_ideal} we obtain that individual components $\mathcal K_{i j}$ of composite (block) operator $\mathcal K$ belong to $K(L^2(\partial\Omega_j), L^2(\partial\Omega_i))$. Now we came to the simple calculations: since our operators act in Hilbert spaces, every $\mathcal K_{i j}$ can be approximated (see Fact~C\ref{Compact_finite_rank}) in the operator norm by some sequence of finite-rank operators $(\mathcal K_{i j,n}^{f r})_{n\in\mathbb N}$, thus we have the sequence of (apparently finite-rank) composite operators $\mathcal K^{f r}_n \mathrel{:=} \{\mathcal K^{f r}_{i j, n}\}_{i,j=1}^N$. After algebraic manipulations we get an estimate (with an $N$-dependent but fixed constant~$C$) $\|\mathcal K - \mathcal K^{f r}_n\|_{B(\mathbf L^2)} \le C \max\limits_{i,j} \|\mathcal K_{i j} - \mathcal K_{i j,n}^{f r}\|_{B(L^2(\partial\Omega_j), L^2(\partial\Omega_i))} \xrightarrow[n\to+\infty]{} 0$, thus $\mathcal K\in K(\mathbf L^2)$ by Fact~C\ref{Compact_finite_rank}. Next, $\mathcal K$ and $\mathring{\mathcal K}$ are topologically equivalent (since $\mathcal S^\kappa$ is a topological isomorphism -- see \eqref{S_kappa_bounded_}), thus $\mathring{\mathcal K}\in K(\mathbf H^1)$ by Fact~C\ref{Compact_fact_ideal}. Finally, owing to the Riesz-Fischer theorem (see \cite[Theorem 2.2.1, Propositions 1.4.9, 2.2.1]{HelemskyAMS}) and Proposition~\ref{proposition_matrix_repr}, operator $\mathbb K \colon \pmb l^2\to\pmb l^2$ is isometrically equivalent to $\mathring{\mathcal K}$, from which we conclude, again with the help of Fact~C\ref{Compact_fact_ideal}, that~$\mathbb K\in K(\pmb l^2)$.~\qed

\subsection{Proof of Proposition~\ref{proposition_spectral_radius}}
\label{Proof_appendix_spectrum_proposition}
\noindent
As we already established in the proof of Proposition~\ref{proposition_K_compact}, $\mathcal K$ and $\mathring{\mathcal K}$ are topologically equivalent, and $\mathring{\mathcal K}$ and $\mathbb K$ are isometrically equivalent; but such equivalences must preserve spectra (due to \cite[Proposition~5.1.1]{HelemskyAMS}), thus we arrive at the important conclusion that $\operatorname{Spectrum}(\mathring{\mathcal K}) = \operatorname{Spectrum}(\mathcal K) = \operatorname{Spectrum}(\mathbb K)$. Also, we already established (in Proposition~\ref{proposition_K_compact}) that these operators are compact, thus Fact~C\ref{Compact_fact_spectrum} assures that the spectra are (at most) countable and $\forall\nu\in\operatorname{Spectrum}(\mathring{\mathcal K})\setminus\{0\}$ is an eigenvalue so that $\exists\Vec{\Phi}_{\text{out}}^{\partial\Omega}\in\mathbf H^1$, $\Vec{\Phi}_{\text{out}}^{\partial\Omega}\ne0$, such that $\mathring{\mathcal K}\Vec{\Phi}_{\text{out}}^{\partial\Omega} = \nu \Vec{\Phi}_{\text{out}}^{\partial\Omega}$ (or alternatively, if working with sequences of Fourier coefficients, $\exists\Vec{\Tilde{\mathbb G}}\in\pmb l^2$, $\Vec{\Tilde{\mathbb G}}\ne\mathbf0$, such that $\mathbb K \Vec{\Tilde{\mathbb G}} = \nu \Vec{\Tilde{\mathbb G}}$); this relation, with the help of \eqref{limiting_values_of_Phi_out_at_boundary_} and \eqref{mathcal_K_Sobolev_representation}, leads to the following relations on the coefficients $\Tilde G_{n m,i}$ of~$\left.\Phi_{\text{out},i}\right|_{\partial\Omega_i}$:
\begin{equation}
\label{eigen_problem_statement0}
\sum_{j=1,j\ne i}^N \, \sum_{L,M} \frac{\beta_{n m, L M}(\Tilde a_i,\varepsilon_i,\mathbf R_{i j})\Upsilon_{L,j}}{\alpha_n(\Tilde a_i,\varepsilon_i) \Upsilon_{n,i}} \Tilde G_{L M,j} = -\frac{1}{\lambda}\Tilde G_{n m,i}
\end{equation}
for all $i\in\overline{1,\ldots,N}$, $n\ge0$, $-n\le m\le n$, where we also denoted $\nu=-1/\lambda$ for further convenience. Furthermore, one may also check that following the derivations in the proof of Proposition~\ref{proposition_matrix_repr}, expressing the actions $\mathcal A_i^{-1}\mathcal B_{i j}\mathfrak g_j$ of operators $\mathcal A_i^{-1}\mathcal B_{i j}$ through the corresponding Fourier coefficients $\mathfrak g_{n m,j}$, and taking into account relation~\eqref{relation_between_G_and_g}, the associated eigenvalue problem to the operator $\mathcal K$ of \eqref{main_eq_G_op_L2} (acting in $\mathbf L^2$) also leads to the same relations~\eqref{eigen_problem_statement0}; the corresponding eigendensities $\mathfrak g_i$ so obtained then give rise (see \eqref{vol_potentials_via_operators}, \eqref{S_vol_definition}) to layer potentials formally defined everywhere in~$\mathbb R^3$ (for them, in order not to clutter up the notations, we will retain the old designations $\Phi_{\text{out},i}$ $\forall i\in\overline{1,\ldots,N}$ and denote $\Phi_{\text{out}} \mathrel{:=} \sum\nolimits_{i=1}^N \Phi_{\text{out},i}$ as usual).

Let $B_R\subset\mathbb R^3$ be a ball of sufficiently large radius $R$ so that it encloses all particles $\Omega_i$ (with no loss of generality one may assume $B_R$ to be centered at e.g.~the origin of a global coordinate system). Employing the Green first identity and PBE (see~\eqref{Lin_eqs_lpb}) we arrive~at
\begin{align*}
0 & \le \kappa^2\int_{\Omega_\text{sol}\cap B_R}\left|\Phi_{\text{out}}\right|^2 d\mathbf r + \int_{\Omega_\text{sol}\cap B_R}\left|\nabla\Phi_\text{out}\right|^2 d\mathbf r \\
& = \int_{\Omega_\text{sol}\cap B_R}\Phi_{\text{out}}^\star \Delta\Phi_\text{out}d\mathbf r + \int_{\Omega_\text{sol}\cap B_R}\left|\nabla\Phi_\text{out}\right|^2 d\mathbf r = \\
& = - \sum_{i=1}^N\oint_{\partial\Omega_i} \Phi_{\text{out}}^\star\frac{\partial\Phi_\text{out}}{\partial\mathbf n_i^+}d s + \oint_{\partial B_R}\Phi_{\text{out}}^\star\frac{\partial\Phi_\text{out}}{\partial\mathbf n^-}d s,
\end{align*}
then the last integral vanishes (due to the asymptotics of $\Phi_\text{out}$, see~\eqref{Lin_eqs_Phi_out}) as $R\to+\infty$, so that we obtain
\begin{equation}
\label{temp_Green_1st}
\begin{aligned}
0 & \le \kappa^2\int_{\Omega_\text{sol}}\left|\Phi_{\text{out}}\right|^2 d\mathbf r + \int_{\Omega_\text{sol}}\left|\nabla\Phi_\text{out}\right|^2 d\mathbf r \\
& = - \sum_{i=1}^N\oint_{\partial\Omega_i} \Phi_{\text{out}}^\star\frac{\partial\Phi_\text{out}}{\partial\mathbf n_i^+}d s .
\end{aligned}
\end{equation}
Note that the use of Green's identity was allowed -- indeed, following the proofs of Lemmas~\ref{Ai_has_bounded_inverse} and \ref{Bij_is_compact} one can derive that operator $\mathring{\mathcal K}\in B(\mathbf H^s,\mathbf H^{s+s_0})$ with an $s_0>0$ independent of $s$ (i.e.~$\mathring{\mathcal K}$ possesses smoothing properties), which entails (by applying $\mathring{\mathcal K}$ iteratively) that its eigenfunction $\Vec{\Phi}_{\text{out}}^{\partial\Omega}$ corresponding to $\nu\ne0$ should lie in $\mathbf H^s$ with arbitrary $s\ge1$. Using mapping properties of layered potential operators (see \cite[Theorem~3.1.16]{SauterSchwab}, \cite[Theorem~6.13]{McLean_0}) one can derive that $\Phi_\text{out}\in H^2(\Omega_\text{sol})$. Thus, the use of Green's identity for~$\Phi_\text{out}$ was permitted. Now taking into account that on $\partial\Omega_i$ we have
\begin{widetext}
\begin{align*}
&\left.\Phi_{\text{out}}^\star\right|_{r_i\to a_i^+} \frac{\partial\Phi_\text{out}}{\partial\mathbf n_i^+} = \biggl(\mathcal S_i^\kappa \mathfrak g_i + \sum_{j=1, j\ne i}^N \gamma_{\partial\Omega_i}\Breve{\mathcal S}_j^\kappa \mathfrak g_j \biggr)^{\!\!\star} \biggl(\Bigl(\frac{1}{2}\mathcal I_i + \mathcal K_i^\kappa{}^\star\Bigr) \mathfrak g_i + \sum_{j=1,\,j\ne i}^N \frac{\partial}{\partial\mathbf n_i^+}\Breve{\mathcal S}_j^\kappa \mathfrak g_j \biggr) = \left|\text{see the proof of Proposition~\ref{proposition_matrix_repr}}\right| \\
& = \biggl( { } \sum_{n,m} \Tilde G_{n m,i} \frac{Y_n^m(\Hat{\mathbf r}_i)}{(2 n+1) a_i} + \sum_{j=1,j\ne i}^N { } \sum_{L,M} \frac{\Tilde G_{L M,j}}{(2 L+1) k_L(\Tilde a_j) a_j} \sum_{n,m} i_n(\Tilde a_i) \mathcal H_{n m}^{L M}(\mathbf{R}_{i j}) Y_n^m(\Hat{\mathbf r}_i) \biggr)^{\!\!\star}\kappa \\
&\ \times \biggl( { } \sum_{n,m}\Bigl(\frac{n}{\Tilde a_i}k_n(\Tilde a_i)-k_{n+1}(\Tilde a_i)\Bigr) \frac{\Tilde G_{n m,i} Y_n^m(\Hat{\mathbf r}_i)}{(2 n+1) k_n(\Tilde a_i) a_i} + \sum_{j=1,j\ne i}^N { } \sum_{L,M} \frac{\Tilde G_{L M,j}}{(2 L+1) k_L(\Tilde a_j) a_j} \sum_{n,m} \Bigl(\frac{n}{\Tilde a_i}i_n(\Tilde a_i)+i_{n+1}(\Tilde a_i)\Bigr) \mathcal H_{n m}^{L M}(\mathbf{R}_{i j}) Y_n^m(\Hat{\mathbf r}_i)\!\biggr)\!,
\end{align*}
substituting the last equality into the right-hand side of \eqref{temp_Green_1st} and using the orthogonality of $\{Y_n^m(\Hat{\mathbf r}_i)\}_{n m}$ and relation \eqref{eigen_problem_statement0} we then arrive at the expression for $ - \sum_{i=1}^N\oint_{\partial\Omega_i} \Phi_{\text{out}}^\star\frac{\partial\Phi_\text{out}}{\partial\mathbf n_i^+}d s\,$:
\begin{align*}
& -\kappa\sum_{i=1}^N a_i^2 \sum_{n,m}\Biggl(\! \frac{\Tilde G_{n m,i}^\star}{(2 n+1) a_i} - \frac{i_n(\Tilde a_i)}{\lambda^{\!\star}}\frac{n (\varepsilon_i-\varepsilon_\text{sol}) k_n(\Tilde a_i)+\Tilde a_i \varepsilon_\text{sol} k_{n+1}(\Tilde a_i)}{n (\varepsilon_i-\varepsilon_\text{sol}) i_n(\Tilde a_i)-\Tilde a_i \varepsilon_\text{sol} i_{n+1}(\Tilde a_i)}\frac{\Tilde G_{n m,i}^\star}{(2 n+1) k_n(\Tilde a_i) a_i} \!\Biggr)\!\Biggl(\! \Bigl(\frac{n}{\Tilde a_i}k_n(\Tilde a_i)-k_{n+1}(\Tilde a_i)\Bigr) \frac{\Tilde G_{n m,i}}{(2 n+1) k_n(\Tilde a_i) a_i} \\
&\quad + \Bigl(\frac{n}{\Tilde a_i}i_n(\Tilde a_i)+i_{n+1}(\Tilde a_i)\Bigr)\frac{-1}{\lambda}  \frac{n (\varepsilon_i-\varepsilon_\text{sol}) k_n(\Tilde a_i)+\Tilde a_i \varepsilon_\text{sol} k_{n+1}(\Tilde a_i)}{n (\varepsilon_i-\varepsilon_\text{sol}) i_n(\Tilde a_i)-\Tilde a_i \varepsilon_\text{sol} i_{n+1}(\Tilde a_i)}\frac{\Tilde G_{n m,i}}{(2 n+1) k_n(\Tilde a_i) a_i} \Biggr) = - \sum_{i=1}^N \sum_{n,m} \frac{|\Tilde G_{n m,i}|^2}{(2 n+1)^2 k_n^2(\Tilde a_i) a_i |\lambda|^2 \mathtt A_{n,i}^2} \\
&\quad \times \Bigl(|\lambda|^2 \mathtt A_{n,i}^2 k_n(\Tilde a_i)\mathtt D_{n,i} -\lambda^{\!\star}\mathtt A_{n,i}\mathtt B_{n,i} k_n(\Tilde a_i)\mathtt C_{n,i} -\lambda \mathtt A_{n,i}\mathtt B_{n,i} i_n(\Tilde a_i)\mathtt D_{n,i} + \mathtt B_{n,i}^2 i_n(\Tilde a_i)\mathtt C_{n,i} \Bigr),
\end{align*}
where 
\begin{align*}
\mathtt A_{n,i} & \mathrel{:=} n (\varepsilon_i-\varepsilon_\text{sol}) i_n(\Tilde a_i)-\Tilde a_i \varepsilon_\text{sol} i_{n+1}(\Tilde a_i), & \mathtt B_{n,i} & \mathrel{:=} n (\varepsilon_i-\varepsilon_\text{sol}) k_n(\Tilde a_i)+\Tilde a_i \varepsilon_\text{sol} k_{n+1}(\Tilde a_i), \\
\mathtt C_{n,i} & \mathrel{:=} n i_n(\Tilde a_i)+\Tilde a_i i_{n+1}(\Tilde a_i), & \mathtt D_{n,i} & \mathrel{:=} n k_n(\Tilde a_i)-\Tilde a_i k_{n+1}(\Tilde a_i).
\end{align*}
However, using \eqref{wronsky_in_kn} we obtain 
\begin{align*}
& -\lambda^{\!\star}\mathtt A_{n,i}\mathtt B_{n,i} k_n(\Tilde a_i)\mathtt C_{n,i} -\lambda \mathtt A_{n,i}\mathtt B_{n,i} i_n(\Tilde a_i)\mathtt D_{n,i} = -\mathtt A_{n,i}\mathtt B_{n,i} \Bigl[ \bigl(2 n k_n(\Tilde a_i) i_n(\Tilde a_i) + (k_n(\Tilde a_i) i_{n+1}(\Tilde a_i) - i_n(\Tilde a_i) k_{n+1}(\Tilde a_i))\Tilde a_i \bigr) \operatorname{Re} \lambda - \frac{\imath}{\Tilde a_i}\operatorname{Im} \lambda \Bigr] \\
&\quad = -\mathtt A_{n,i}\mathtt B_{n,i} \Bigl[ \digamma_{n,i}\Tilde a_i i_n(\Tilde a_i) k_n(\Tilde a_i) \operatorname{Re} \lambda - \frac{\imath}{\Tilde a_i}\operatorname{Im} \lambda \Bigr], \quad \text{where} \quad \digamma_{n,i} \mathrel{:=} \frac{2 n}{\Tilde a_i} + \frac{i_{n+1}(\Tilde a_i)}{i_{n}(\Tilde a_i)} - \frac{k_{n+1}(\Tilde a_i)}{k_{n}(\Tilde a_i)};
\end{align*}
from this one concludes that $\operatorname{Im} \lambda$ must be zero, otherwise it contradicts to the fact that $ - \sum_{i=1}^N\oint_{\partial\Omega_i} \Phi_{\text{out}}^\star\frac{\partial\Phi_\text{out}}{\partial\mathbf n_i^+}d s$ is real (see the left-hand side of~\eqref{temp_Green_1st}). From this and \eqref{temp_Green_1st} we thus have
\begin{equation}
\label{temp_Green_1st_1}
0 \le \kappa^2\int_{\Omega_\text{sol}}\left|\Phi_{\text{out}}\right|^2 d\mathbf r + \int_{\Omega_\text{sol}}\left|\nabla\Phi_\text{out}\right|^2 d\mathbf r = - \sum_{i=1}^N\oint_{\partial\Omega_i} \Phi_{\text{out}}^\star\frac{\partial\Phi_\text{out}}{\partial\mathbf n_i^+}d s =
- \sum_{i=1}^N \sum_{n,m} \frac{|\Tilde G_{n m,i}|^2 f_{n,i}(\lambda) }{(2 n+1)^2 k_n^2(\Tilde a_i) a_i \lambda^2 \mathtt A_{n,i}^2} ,
\end{equation}
where the quadratic (with respect to $\lambda$) function $$f_{n,i}(\lambda) \mathrel{:=} \mathtt A_{n,i}^2 \mathtt D_{n,i} k_n(\Tilde a_i) \lambda^2 - \mathtt A_{n,i}\mathtt B_{n,i} \Tilde a_i i_n(\Tilde a_i) k_n(\Tilde a_i) \digamma_{n ,i} \lambda + \mathtt B_{n,i}^2 \mathtt C_{n,i} i_n(\Tilde a_i) .$$ 
\end{widetext}
Now the situation boils down to studying the sign of function $f_{n,i}(\lambda)$ depending on~$\lambda$. Namely, let us show that $f_{n,i}(\lambda)\ge0$ for arbitrary $\lambda\in[-1, 1]$, which will, in turn, entail that there is no nonzero eigenfunctions of $\mathring{\mathcal K}$ (respectively, eigenvectors of $\mathbb K$) as $|\nu|\ge1$, thus such values of $\nu$ can not be eigenvalues. Namely, provided that $\varepsilon_i \le \varepsilon_\text{sol}$ (and this is the only place of proofs, where this assumption is used), we will show that $f_{n,i}(\lambda)>0$ as $n\ge 1$, $\lambda\in[-1, 1]$, while $f_{0,i}(\lambda)>0$ as $\lambda\in[-1, 1)$, $f_{0,i}(1)=0$ (later we will pay particular attention to this case). Indeed, since $f_{n,i}(\lambda)$ is a quadratic parabola with respect to $\lambda$ and with a strictly negative leading coefficient owing to $\mathtt D_{n,i} = \Tilde a_i k_n'(\Tilde a_i) < 0$ (see \eqref{diff_modifiedBessel} and take into account that $k_n(x)$ is monotonically decreasing function in $x>0$) and $\mathtt A_{n,i}<0$ (see \eqref{in_kn_are_positive_} and take into account our assumption $\varepsilon_i \le \varepsilon_\text{sol}$), hence it is sufficient to study the sign of $f_{n,i}(\lambda)$ at points $\lambda = -1,\, 0,\, 1$. After simple but bulky algebraic transformations, using \eqref{wronsky_in_kn} we arrive at $$f_{n,i}(1) = \frac{n \varepsilon_i \varepsilon_\text{sol}}{\Tilde a_i^2}\ge0, \qquad f_{n,i}(0) = \mathtt B_{n,i}^2 \mathtt C_{n,i} i_n(\Tilde a_i) >0,$$ the last inequality is due to $\mathtt B_{n,i} = \alpha_n(\Tilde a_i,\varepsilon_i)\Tilde a_i > 0$ (note that $\alpha_n(\Tilde a_i,\varepsilon_i)>0$ for $\forall n\ge0$) and $\mathtt C_{n,i} i_n(\Tilde a_i) > 0 $ (see~\eqref{in_kn_are_positive_}). Concerning $f_{n,i}(-1)$, let us observe that $$f_{n,i}(-1) = f_{n,i}(1) + 2\mathtt A_{n,i}\mathtt B_{n,i} \Tilde a_i i_n(\Tilde a_i) k_n(\Tilde a_i) \digamma_{n ,i},$$ whence taking into account previously established inequalities $\mathtt B_{n,i}>0$, $\mathtt A_{n,i}<0$, and $\Tilde a_i i_n(\Tilde a_i) k_n(\Tilde a_i)>0$ (see~\eqref{in_kn_are_positive_}), we will immediately obtain the desired inequality $f_{n,i}(-1)>0$ if we show that $\digamma_{n ,i}<0$. But the last inequality can be proved by the following technical lemma.
\begin{lemma}
\label{digamma_n_i_is_negative}
For arbitrary $n\ge0$ and $\Tilde a_i>0$ it holds $\digamma_{n ,i}<0$.
\end{lemma}
\emph{Proof.} We use the following fine inequality\footnote{Let us point out the nontriviality of estimates \eqref{Segura2011_inequality} -- while the left-hand side inequality of \eqref{Segura2011_inequality} was previously known in the literature (at least, for $\nu\ge1/2$, see \cite{Amos1974}), the right-hand side estimate seems to appear for the first time in \cite{Segura_jmaa_2011} and is based on subtle analysis of the associated Riccati differential equations.} proved in recent~\cite[Theorem~1]{Segura_jmaa_2011} and valid for arbitrary $\nu>0$ and $x>0$:
\begin{equation}
\frac{I_{\nu+1/2}(x)}{I_{\nu-1/2}(x)}<\frac{x}{\nu+\sqrt{\nu^2+x^2}}<\frac{K_{\nu-1/2}(x)}{K_{\nu+1/2}(x)}
.\label{Segura2011_inequality}
\end{equation}
Thence one has for~$n\ge1$:
\begin{align*}
& \digamma_{n ,i} = \frac{2 n}{\Tilde a_i} + \frac{i_{n+1}(\Tilde a_i)}{i_{n}(\Tilde a_i)} - \frac{k_{n+1}(\Tilde a_i)}{k_{n}(\Tilde a_i)} = \left|\text{use \eqref{in_kn_recurrences0}, \eqref{in_kn_are_positive_}}\right| = \frac{2 n}{\Tilde a_i} \\ 
& \  + \frac{i_{n+1}(\Tilde a_i)}{i_n(\Tilde a_i)} - \frac{k_{n-1}(\Tilde a_i)}{k_n(\Tilde a_i)} - \frac{2 n+1}{\Tilde a_i} = -\frac{1}{\Tilde a_i} + \frac{i_{n+1}(\Tilde a_i)}{i_n(\Tilde a_i)} \\ 
& \  - \frac{k_{n-1}(\Tilde a_i)}{k_n(\Tilde a_i)} < \left|\text{use~\eqref{Segura2011_inequality}}\right| < -\frac{1}{\Tilde a_i} + \frac{\Tilde a_i}{n+1+\sqrt{\Tilde a_i^2 + (n+1)^2}} \\ 
& \  - \frac{\Tilde a_i}{n+\sqrt{\Tilde a_i^2 + n^2}} < 0.
\end{align*}
Finally, for $n=0$ one directly obtains $\digamma_{0 ,i} = (\Tilde a_i \coth\Tilde a_i - \Tilde a_i -2)/\Tilde a_i < 0$.~\qed

\emph{Continuation of the proof of Proposition~\ref{proposition_spectral_radius}.} Lemma~\ref{digamma_n_i_is_negative} together with the above estimates show that $f_{n,i}(\lambda)\ge0$ for all $|\lambda|\le1$ and the equality (i.e.~$f_{n,i}(\lambda)=0$) takes place only when $n=0$, $\lambda=1$. Thus, we come to the situation when the left-hand side of~\eqref{temp_Green_1st_1}, i.e.~$\kappa^2\int_{\Omega_\text{sol}}\left|\Phi_{\text{out}}\right|^2 d\mathbf r + \int_{\Omega_\text{sol}}\left|\nabla\Phi_\text{out}\right|^2 d\mathbf r$, is $\ge0$, while the right-hand side, that is $- \sum_{i=1}^N \sum_{n,m} \frac{|\Tilde G_{n m,i}|^2 f_{n,i}(\lambda) }{(2 n+1)^2 k_n^2(\Tilde a_i) a_i \lambda^2 \mathtt A_{n,i}^2} $, is $\le0$: but this can only be realized when both these expressions are simultaneously equal to zero, thus $\Phi_\text{out}(\mathbf r)=0$ $\forall \mathbf r\in\Omega_\text{sol}$ as well as all $\Tilde G_{n m,i}$ should be equal to zero, which hereby confirm that there are no eigenvalues $\nu$ such that $|\nu|\ge1$. The only subtle moment here to be additionally clarified is that one may conjecture that $\Tilde G_{0 0,i} \ne 0$ in the case when $\lambda=1$ -- because, as it was discussed above, one has $f_{0,i}(1)=0$, which then in principle allows for the right-hand side of~\eqref{temp_Green_1st_1} to be zero even if some $\Tilde G_{0 0,i}$ are nonzero. Let us now show, however, that $\Phi_\text{out}(\mathbf r)=0$ $\forall \mathbf r\in\Omega_\text{sol}$ (that comes from the left-hand side of~\eqref{temp_Green_1st_1}) prevents this possibility. Indeed, since spheres do not overlap ($R_{i j}>a_i+a_j$), we can consider the potential $0 = \Phi_\text{out}=\sum_{i=1}^N G_{0 0,i}k_0(\Tilde r_i)/\sqrt{4\pi}$ (see \eqref{Lin_eq_tot_pot}, \eqref{Lin_eqs_Phi_out}) and its normal derivative (the external potential is continuously differentiable, see the proof of Lemma~\ref{Bij_is_compact}) at a spherical surface $r_i=b_i>a_i$ close to $\partial\Omega_i$ and lying inside~$\Omega_\text{sol}$; using relation \eqref{addition_theorem_screened_0}, formulas \eqref{diff_modifiedBessel}, and integrating over sphere $r_i=b_i$ we thus arrive at system $k_0(\Tilde b_i) G_{0 0,i} + i_0(\Tilde b_i)\sum_{j=1,j\ne i}^N k_0(\Tilde R_{i j}) G_{0 0,j} = 0$,  $-k_1(\Tilde b_i) G_{0 0,i} + i_1(\Tilde b_i)\sum_{j=1,j\ne i}^N k_0(\Tilde R_{i j}) G_{0 0,j} = 0$, from which we conclude that $G_{0 0,i} = \sum_{j=1,j\ne i}^N k_0(\Tilde R_{i j}) G_{0 0,j} = 0$ because $\det\left(\begin{smallmatrix}k_0(\Tilde b_i) & i_0(\Tilde b_i)\\ -k_1(\Tilde b_i) & i_1(\Tilde b_i) \end{smallmatrix} \right) =\left|\text{see~\eqref{wronsky_in_kn}}\right|=\Tilde b_i^{-2}\ne0$; such procedure can be repeated for $\forall i\in\overline{1,\ldots,N}$, thus giving all $\Tilde G_{0 0,i} = 0$. These show that there is no nontrivial (nonzero) eigenfunction (eigenvector) corresponding to $|\nu|\ge1$, thus such $\nu$ do not belong to the spectrum. Then Fact~C\ref{Compact_fact_spectrum} completely witnesses the validity of Proposition~\ref{proposition_spectral_radius}.~\qed
\begin{remark}
\emph{(On an extension of the proof of Proposition~\ref{proposition_spectral_radius} to the case $\kappa=0$.)} A detailed formal proof of this case is beyond the scope of this paper, however let us note that in the Poisson limit, i.e.~at zero ionic strength as $\kappa\to0$, the left-hand side of \eqref{temp_Green_1st_1} will yield $\int_{\Omega_\text{sol}}\left|\nabla\Phi_\text{out}\right|^2 d\mathbf r = 0$ in the proposed scheme of the proof, which together with the condition $\Phi_\text{out}(\infty)=0$ still provides the key equality~$\Phi_\text{out}=0$ employed in the current proof. Furthermore, since we are using the scaled distances (see $\Tilde r_i$ in representation \eqref{Lin_eqs_Phi_in_out}) as $\kappa>0$, in order to handle the limit case $\kappa=0$ the expansion coefficients must also be properly rescaled (see \cite[Appendix~F]{our_jpcb} where two-body cases were discussed) by taking into account the small-argument asymptotics \eqref{small_Bessel_i_k} of Bessel functions.
\end{remark}
\begin{remark}
\label{Remark_on_proof_epsiloni_less_epsilonm}
\emph{(On the asymptotics of matrix elements of operator~$\mathbb K$.)} It is shown in the joint paper \textcolor{red}{\cite{supplem_pre}} that $\frac{\beta_{n m, L M}(\Tilde a_i,\varepsilon_i,\mathbf R_{i j})}{\alpha_n(\Tilde a_i,\varepsilon_i)} = O(\Tilde a_i^{\max(3,\, 2 n+1)})$ and $\frac{\beta_{n m, L M}(\Tilde a_i,\varepsilon_i,\mathbf R_{i j})\Upsilon_{L,j}}{\alpha_n(\Tilde a_i,\varepsilon_i)\Upsilon_{n,i}} = O(\Tilde a_i^{\max(3,\, 2 n+1)-n})$ for small $\Tilde a_i\to0$, while for large $\Tilde R_{i j}\to+\infty$ it holds $\frac{\beta_{n m, L M}(\Tilde a_i,\varepsilon_i,\mathbf R_{i j})}{\alpha_n(\Tilde a_i,\varepsilon_i)} = O(e^{-\Tilde R_{i j}}/\Tilde R_{i j})$ and $\frac{\beta_{n m, L M}(\Tilde a_i,\varepsilon_i,\mathbf R_{i j})\Upsilon_{L,j}}{\alpha_n(\Tilde a_i,\varepsilon_i)\Upsilon_{n,i}} = O(e^{-\Tilde R_{i j}}/\Tilde R_{i j})$; these asymptotics are valid regardless the condition $\varepsilon_i \le \varepsilon_\text{sol}$ used in the proof of Proposition~\ref{proposition_spectral_radius}. Hence, (dimensionless) elements of each block $\mathsf A_i^{-1}\mathsf B_{i j}$ of $\mathbb K$ anyway decay as particles' radii decrease or $R_{i j}$ values increase, so that one then may anticipate that $\mathbb K$ becomes a small perturbation of $\mathbb I$ in \eqref{global_lin_sys1} as particles shrink or go away from each other, which in turn would justify the statement of Proposition~\ref{proposition_spectral_radius} (and thence the operator inversion $(\mathbb I + \mathbb K)^{-1} = \sum\nolimits_{\ell=0}^{+\infty}(-1)^\ell \mathbb K^\ell$ crucial for constructing the screening-ranged expansions in~\textcolor{red}{\cite{supplem_prl,supplem_pre,supplem_pre_force}}).
\end{remark}

\subsection{Proofs of Corollaries \ref{corollary_Neumann_covergence} and \ref{corollary_solution_exist_unique}}
\label{Proof_appendix_spectrum_corollaries_converg}
\noindent
Proposition~\ref{proposition_spectral_radius} shows that for spectral radius we have $r(\mathring{\mathcal K}) < 1$, from which now the convergence of Neumann series $\sum_{k=0}^{+\infty}(-1)^k\mathring{\mathcal K}^k$ to the operator $(\mathcal I + \mathring{\mathcal K})^{-1}\in B(\mathbf H^1)$ in the operator norm $\|\cdot\|=\|\cdot\|_{B(\mathbf H^1)}$ immediately follows by the standard functional-analytic argumentation \cite{HelemskyAMS} (which we very briefly remind here just for the sake of completeness the presentation). Indeed, $\operatorname{Spectrum}(\mathring{\mathcal K})$ is a compact non-empty set \cite{HelemskyAMS} since $\mathring{\mathcal K}\in B(\mathbf H^1)$ (moreover, $\mathring{\mathcal K}\in K(\mathbf H^1)$ so Fact~C\ref{Compact_fact_spectrum} applies) and the Gelfand formula for spectral radius shall yield $r(\mathring{\mathcal K}) = \lim_{k\to+\infty}\sqrt[k]{\|\mathring{\mathcal K}^k\|}$, thus for an arbitrarily fixed $\delta\in(r(\mathring{\mathcal K}), 1)$ there exists $N_0$ such that $\forall k\ge N_0$ it holds $\|\mathring{\mathcal K}^k\|<\delta^k$: this confirms that the numeric series $\sum_{k=0}^{+\infty}\|\mathring{\mathcal K}^k\|$ converges and thus $\sum_{k=0}^{+\infty}(-1)^k\mathring{\mathcal K}^k$ converges absolutely to some element of $B(\mathbf H^1)$ (because $B(\mathbf H^1)$ is a Banach space, so any absolutely convergent series converges to some element of the space, see \cite[Proposition~2.1.8]{HelemskyAMS}); $(\mathcal I + \mathring{\mathcal K})\lim_{k\to+\infty}(1- \mathring{\mathcal K} + \cdots +(-1)^k\mathring{\mathcal K}^k) =\bigl|$mapping $x\mapsto A\cdot x$ is continuous in $B(\mathbf H^1)$ $\forall A\in B(\mathbf H^1)$, thus $\mathcal I + \mathring{\mathcal K}$ can be moved inside the limit$\bigr| =\lim_{k\to+\infty}(\mathcal I + \mathring{\mathcal K})(1- \mathring{\mathcal K} + \cdots +(-1)^k\mathring{\mathcal K}^k) = \lim_{k\to+\infty}(\mathcal I - (-1)^{k+1} \mathring{\mathcal K}^{k+1}) = \mathcal I = \left|\text{essentially similar calculations}\right| = (\lim_{k\to+\infty}(1- \mathring{\mathcal K} + \cdots +(-1)^k\mathring{\mathcal K}^k)) (\mathcal I + \mathring{\mathcal K})$, thus $\exists(\mathcal I + \mathring{\mathcal K})^{-1}\in B(\mathbf H^1)$ and the series $\sum_{k=0}^{+\infty}(-1)^k\mathring{\mathcal K}^k$ converges exactly to it. These considerations are transferred without changes to the accompanying series $\sum_{k=0}^{+\infty}(-1)^k\mathcal K^k$ and $\sum_{k=0}^{+\infty}(-1)^k\mathbb K^k$ absolutely converging in the associated operator norms $\|\cdot\|_{B(\mathbf L^2)}$ and $\|\cdot\|_{B(\pmb l^2)}$, respectively.~\qed
\begin{remark}
Let us note that for any finite sequence $\mathfrak X_1,\ldots,\mathfrak X_N$ of Banach spaces there is a canonical bijective correspondence between operators from $B(\bigoplus_{i=1}^N\mathfrak X_i)$ and $N\times N$-matrices composed of individual operators from $B(\mathfrak X_j,\mathfrak X_i)$ ($i,j\in\overline{1,\ldots,N}$); this correspondence is linear and \emph{multiplicative} (rigorously speaking, it delivers a bijective algebraic homomorphism completely identifying the Banach algebra $B(\bigoplus_{i=1}^N\mathfrak X_i)$ with the set of such composite block matrices, see \cite{Lautsen2001}). This justifies the possibility of matrix-like multiplication operations (particularly important for the expansion theory developed in the joint papers \textcolor{red}{\cite{supplem_prl,supplem_pre,supplem_pre_force}}) when calculating powers (see Corollary~\ref{corollary_Neumann_covergence}) of the composite block NP-type operators introduced in the current study.
\end{remark}
\subsection{Proof of Corollary~\ref{corollary_solutions_real}}
\label{appendix_proof_corollary_solutions_real}
\noindent
Since potential \eqref{varPhi_in_i_multipoles} is real, components $S_{n m,i}$ and $E_{n m,i}$ of the right-hand side vectors $\mathbf S_i = \{S_{n m,i}\}_{n m}$ (see \eqref{eqs_G_intermediate_}, \eqref{various_matrix_definitions_0}) and $\mathbf E_i = \{E_{n m,i}\}_{n m}$ (the right-hand side vector in the representation $\mathbf L_i^{(0)} = \mathsf C_i \Tilde{\mathbf G}_i^{(0)} + \mathbf E_i$ constructed in \textcolor{red}{\cite{supplem_prl,supplem_pre}}) satisfy relations $S_{n, -m,i} = (-1)^m S_{n m,i}^\star$ and $E_{n, -m,i} = (-1)^m E_{n m,i}^\star$ (superscript $\star$ stands here for the complex conjugation) for arbitrary indices $0\le|m|\le n$ and $1\le i\le N$ (note also that relation $S_{n, -m,i} = (-1)^m S_{n m,i}^\star$ holds also in the case of presence the real-valued surface free charge distributions $\sigma_i^\text{f}$ -- see~\eqref{F_for_surface_charge_vec_}). 

In the following paragraphs we will leverage the componentwise representations of screening-ranged vector addends $\mathbf L_i^{(\ell)}$ and $\Tilde{\mathbf G}_i^{(\ell)}$ written out in~\textcolor{red}{\cite{supplem_pre}} in details.

\emph{Potential \eqref{Lin_eqs_Phi_out} is real.} To prove this, we must show that its expansion coefficients $\{G_{n m,i}\}_{n m}$ fulfill $G_{n,-m,i}=(-1)^m G_{n m,i}^\star$. To do so, we employ expansion $\Tilde{\mathbf G}_i = \sum_{\ell=0}^{+\infty}\Tilde{\mathbf G}_i^{(\ell)}$ in ascending screening orders resulting from the Neumann type series expansion built in \textcolor{red}{\cite{supplem_prl,supplem_pre}}. Indeed, from the componentwise expression $\Tilde G_{n m,i}^{(0)} = \frac{S_{n m,i}}{\alpha_n(\Tilde a_i,\varepsilon_i) \Upsilon_{n,i}}$ (see \textcolor{red}{\cite{supplem_pre}}) we immediately see that $\Tilde G_{n,-m,i}^{(0)}=(-1)^m \Tilde G_{n m,i}^{(0) \star}$ holds. Now we use mathematical induction on index $\ell$: assuming that the equality $\Tilde G_{n,-m,i}^{(\ell-1)}=(-1)^m \Tilde G_{n m,i}^{(\ell-1) \star}$ is true we prove its validity for the next index value, that is $\Tilde G_{n,-m,i}^{(\ell)}=(-1)^m \Tilde G_{n m,i}^{(\ell) \star}$. Indeed, from equality $\Tilde{\mathbf G}_i^{(\ell)} = -\sum\nolimits_{j=1,\, j\ne i}^N \mathsf A_i^{-1}\mathsf B_{i j} \Tilde{\mathbf G}_j^{(\ell-1)}$ and relation $\beta_{n, -m, L M}(\Tilde a_i,\varepsilon_i,\mathbf R_{i j}) = (-1)^{m-M} \beta_{n m, L, -M}(\Tilde a_i,\varepsilon_i,\mathbf R_{i j})^\star$, which easily follows from \eqref{beta_definition} and the symmetry 
\begin{equation}
\label{symmetry_H_nmLM_1}
\mathcal H_{n, -m}^{L, -M}(\mathbf{R}_{i j}) = (-1)^{m+M}\mathcal H_{n m}^{L M}(\mathbf{R}_{i j})^\star
\end{equation}
established in \textcolor{red}{\cite{supplem_pre}}, we obtain
\begin{align*}
& \Tilde G_{n,-m,i}^{(\ell)} = \sum_{j=1,\,j\ne i}^N\sum_{L,M}\frac{\beta_{n, -m, L M}(\Tilde a_i,\varepsilon_i,\mathbf R_{i j})\Upsilon_{L,j}}{\alpha_n(\Tilde a_i,\varepsilon_i) \Upsilon_{n,i}}\Tilde G_{L M,j}^{(\ell-1)} \\
& = \sum_{j=1,\,j\ne i}^N\sum_{L,M}\frac{(-1)^{m-M}\beta_{n m, L, -M}(\Tilde a_i,\varepsilon_i,\mathbf R_{i j})^\star}{\alpha_n(\Tilde a_i,\varepsilon_i) \Upsilon_{n,i} \Upsilon_{L,j}^{-1}}(-1)^M \Tilde G_{L, -M,j}^{(\ell-1) \star} \\
& = \left|\text{change summation index $M\to-M$ (here $-L\le M\le L$)}\right| \\
& = (-1)^m \Tilde G_{n m,i}^{(\ell) \star} .
\end{align*}
Thus, we have relations $\Tilde G_{n,-m,i}^{(\ell)}=(-1)^m \Tilde G_{n m,i}^{(\ell) \star}$ valid $\forall\ell\ge0$, that entail relation $G_{n,-m,i}=(-1)^m G_{n m,i}^\star$ also for the coefficients $\{G_{n m,i}\}_{n m}$ of potential~\eqref{Lin_eqs_Phi_out}.

\emph{Potential \eqref{Lin_eqs_Phi_in} is real.} From the representation $L^{(\ell)}_{n m,i} = \sum\limits_{j=1,\, j\ne i}^N \; \sum\limits_{L,M} \frac{\varepsilon_\text{sol} \mathcal H_{n m}^{L M}(\mathbf R_{i j}) \Upsilon_{L,j}}{\Tilde a_i^{n+2} \alpha_n(\Tilde a_i,\varepsilon_i)}\Tilde G^{(\ell-1)}_{L M,j}$ (see \textcolor{red}{\cite{supplem_pre}}) we have $\forall\ell\ge1$:
\begin{align*}
L^{(\ell)}_{n, -m,i} & = \sum_{j=1,\, j\ne i}^N \sum_{L,M} \frac{\varepsilon_\text{sol} \mathcal H_{n, -m}^{L M}(\mathbf R_{i j}) \Upsilon_{L,j}}{\Tilde a_i^{n+2} \alpha_n(\Tilde a_i,\varepsilon_i)}\Tilde G^{(\ell-1)}_{L M,j} \\
& =\Bigl/\text{use $\mathcal H_{n, -m}^{L M}(\mathbf{R}_{i j}) = (-1)^{m-M}\mathcal H_{n m}^{L, -M}(\mathbf{R}_{i j})^\star$} \\
& \quad\ \text{(see~\eqref{symmetry_H_nmLM_1}) and $\Tilde G^{(\ell-1)}_{L M,j}=(-1)^M \Tilde G_{L, -M,j}^{(\ell-1) \star}$,} \\
& \quad\ \text{then change summation index $M\to-M$}\Bigr/ \\
& = (-1)^m L^{(\ell) \star}_{n m,i}\, ;
\end{align*}
as well, from $L^{(0)}_{n m,i} = \frac{k_n(\Tilde a_i) S_{n m,i}}{\Tilde a_i^n \alpha_n(\Tilde a_i,\varepsilon_i)} + E_{n m,i}$ (see \textcolor{red}{\cite{supplem_pre}}) we have $L^{(0)}_{n, -m,i} = (-1)^m L^{(0) \star}_{n m,i}$. Hence, the expansion coefficients $L_{n m,i}=\sum_{\ell=0}^{+\infty}L_{n m,i}^{(\ell)}$ of \eqref{Lin_eqs_Phi_in} obey relation $L_{n,-m,i}=(-1)^m L_{n m,i}^\star$.~\qed

\subsection{On the convergence of series \eqref{scr_ranged_expansion_G}-\eqref{energy_expansion_components_abs_gen}}
\label{appendix_scr_rang_expansions_G_L_converg}
\noindent
Let us first show that the numerical series \eqref{scr_ranged_expansion_G} and \eqref{scr_ranged_expansion_L} converge absolutely. Indeed, based on the construction of series \eqref{scr_ranged_expansion_G} we easily obtain estimates
\begin{equation}
\label{temp_eq_G_ell_converge_l2}
\begin{aligned}
& \sum_{\ell=0}^{+\infty}|\Tilde G_{n m,i}^{(\ell)}| \le \sum_{\ell=0}^{+\infty}\sqrt{\sum_{n,m}|\Tilde G_{n m,i}^{(\ell)}|^2} \le \sum_{\ell=0}^{+\infty}\|\Vec{\Tilde{\mathbb G}}^{(\ell)}\|_{\pmb l^2} \\
&\quad \le \Bigl( \, \sum_{\ell=0}^{+\infty} \|\mathbb K^\ell\|_{B(\pmb l^2)} \Bigr)\|\mathbb A^{-1} \Vec{\mathbb S}\|_{\pmb l^2}<\infty
\end{aligned}
\end{equation}
since the series $\sum\nolimits_{\ell=0}^{+\infty} \|\mathbb K^\ell\|_{B(\pmb l^2)}$ converges (see comments in Sec.~\ref{Proof_appendix_spectrum_corollaries_converg}), which confirms the absolute convergence of~\eqref{scr_ranged_expansion_G}. As for the series \eqref{scr_ranged_expansion_L}, its addends are assembled from relation \eqref{implement_bc1}, i.e.~$L_{n m,i} = - \frac{\Hat L_{n m,i}}{\Tilde a_i^{2 n+1}} + \frac{k_n(\Tilde a_i)}{\Tilde a^n_i} \Upsilon_{n,i} \Tilde G_{n m,i} + \frac{i_n(\Tilde a_i)}{\Tilde a^n_i}\sum\limits_{j=1,\, j\ne i}^N \, \sum\limits_{L,M}\mathcal H_{n m}^{L M}(\mathbf R_{i j}) \Upsilon_{L,j} \Tilde G_{L M,j}$ (taking into account also scaling \eqref{G_nm_scaling}), by substituting into it the corresponding expansions \eqref{scr_ranged_expansion_G} for $\Tilde G_{n m,i}$ and~$\Tilde G_{L M,j}$:
\begin{widetext}
\begin{subequations}
\label{scr_ranged_expansion_L_ell_explicit}
\begin{align}
L_{n m,i} & = - \frac{\Hat L_{n m,i}}{\Tilde a_i^{2 n+1}} + \frac{k_n(\Tilde a_i)}{\Tilde a^n_i} \Upsilon_{n,i} \sum_{\ell=0}^{+\infty} \Tilde G_{n m,i}^{(\ell)} + \frac{i_n(\Tilde a_i)}{\Tilde a^n_i}\sum_{j=1,\, j\ne i}^N \, \sum_{L,M}\mathcal H_{n m}^{L M}(\mathbf R_{i j}) \Upsilon_{L,j} \sum_{\ell=0}^{+\infty}\Tilde G_{L M,j}^{(\ell)} \label{scr_ranged_expansion_L_ell_explicit_1} \\
& = \underbrace{- \frac{\Hat L_{n m,i}}{\Tilde a_i^{2 n+1}} + \frac{k_n(\Tilde a_i)}{\Tilde a^n_i} \Upsilon_{n,i} \Tilde G_{n m,i}^{(0)}}_{L_{n m,i}^{(0)}} \, + \, \sum_{\ell=1}^{+\infty}\biggl(\underbrace{\frac{k_n(\Tilde a_i)}{\Tilde a^n_i} \Upsilon_{n,i} \Tilde G_{n m,i}^{(\ell)} + \frac{i_n(\Tilde a_i)}{\Tilde a^n_i}\sum_{j=1,\, j\ne i}^N \, \sum_{L,M}\mathcal H_{n m}^{L M}(\mathbf R_{i j}) \Upsilon_{L,j} \Tilde G_{L M,j}^{(\ell-1)}}_{L_{n m,i}^{(\ell)}}\biggr) ; \label{scr_ranged_expansion_L_ell_explicit_2}
\end{align}
\end{subequations}
\end{widetext}
to justify the change of orders of summations (over $\ell$ and $L, M$) when obtaining the last equality, it is sufficient to prove that the multiple series $\sum_{\ell=0}^{+\infty} \sum_{L,M} |\mathcal H_{n m}^{L M}(\mathbf R_{i j}) \Upsilon_{L,j} \Tilde G_{L M,j}^{(\ell)}|$ converges. To do so we obtain
\begin{align}
& \sum_{\ell=0}^{+\infty} \, \sum_{L,M} |\mathcal H_{n m}^{L M}(\mathbf R_{i j}) \Upsilon_{L,j} \Tilde G_{L M,j}^{(\ell)}| \le\left|\text{use Cauchy-Schwarz ineq.}\right| \notag\\ 
& \le \sqrt{\sum_{L,M}|\mathcal H_{n m}^{L M}(\mathbf R_{i j}) \Upsilon_{L,j}|^2} \sum_{\ell=0}^{+\infty}\sqrt{\sum_{L,M}|\Tilde G_{L M,j}^{(\ell)}|^2} < \infty \label{double_series_in_L_abs_convergence}
\end{align}
--- indeed, $\sum\limits_{\ell=0}^{+\infty}\sqrt{\sum\limits_{L,M}|\Tilde G_{L M,j}^{(\ell)}|^2} \le \sum\limits_{\ell=0}^{+\infty}\|\Vec{\Tilde{\mathbb G}}^{(\ell)}\|_{\pmb l^2} < \infty$ (see \eqref{temp_eq_G_ell_converge_l2}), while $\sqrt{\sum\limits_{L,M}|\mathcal H_{n m}^{L M}(\mathbf R_{i j}) \Upsilon_{L,j}|^2}<\infty$ follows from a necessary condition for an infinite-size matrix to describe a bounded operator on a separable Hilbert space\footnote{Namely, the condition $\sup_n\{\sum_m|a_{m n}|^2\}+\sup_m\{\sum_n|a_{m n}|^2\}<\infty$ is necessary for $\{a_{m n}\}$ to be a matrix of a bounded operator, see~\cite[Sec.~2.2]{HelemskyAMS}.\label{footnote_nec_cond_bound_op}} by taking into account that block operator $\mathring{\mathcal M} \mathrel{:=} \{\gamma_{\partial\Omega_i}\Breve{\mathcal S}_j^\kappa (\mathcal S_j^\kappa)^{-1}\}_{i,j=1}^N\in B(\mathbf H^1)$ (see Remark~\ref{remark_the_same_notation_potentials_surface_volumetric} and Sec.~\ref{Proof_appendix_representation_proposition}) and for the matrix representation of its component operator $\gamma_{\partial\Omega_i}\Breve{\mathcal S}_j^\kappa (\mathcal S_j^\kappa)^{-1} \in B(H^1(\partial\Omega_j), H^1(\partial\Omega_i))$ we obtain 
\begin{align}
& \gamma_{\partial\Omega_i}\Breve{\mathcal S}_j^\kappa (\mathcal S_j^\kappa)^{-1}\tfrac{Y_L^M(\Hat{\mathbf r}_j)}{(2 L+1) a_j} \notag \\ 
&\quad = \sum\limits_{n,m} (2 n+1) a_i i_n(\Tilde a_i) \mathcal H_{n m}^{L M}(\mathbf{R}_{i j}) \Upsilon_{L,j} \tfrac{Y_n^m(\Hat{\mathbf r}_i)}{(2 n+1) a_i} \label{matrix_representation_M_for_L}
\end{align}
(here the calculations are rather similar to those of Sec.~\ref{Proof_appendix_representation_proposition}). Now convergences \eqref{temp_eq_G_ell_converge_l2} and \eqref{double_series_in_L_abs_convergence} ensure the possibility of the required rearrangements of the orders of summations when passing from \eqref{scr_ranged_expansion_L_ell_explicit_1} to \eqref{scr_ranged_expansion_L_ell_explicit_2}, as well as the absolute convergence of series~\eqref{scr_ranged_expansion_G}-\eqref{scr_ranged_expansion_L}; based on this, further analysis and transformations of the terms of these series are carried out in our joint study~\textcolor{red}{{\cite{supplem_pre}}}.

The addends of the screening-ranged energy expansion \eqref{energy_expansion_components_abs_gen} have the form~\textcolor{red}{\cite{supplem_pre,supplem_prl}}
\begin{equation}
\label{energy_expansion_components_abs_center}
\mathcal E^{(\ell)} = \frac{1}{2}\sum_{i=1}^N\frac{\varepsilon_i \varepsilon_0}{\kappa}\sum_{n,m}(2 n+1)\Hat L_{n m,i}^\star L_{n m,i}^{(\ell)}
\end{equation}
for the fixed charge inside spheres (i.e.~described by $\rho_i^\text{f}$), or $$\mathcal E^{(\ell)} = \frac{1}{2}\sum_{i=1}^N a_i^2 \sum_{n,m} \sigma_{n m,i}^{\text{f} \ \star} \Tilde a_i^n L_{n m,i}^{(\ell)}$$ when one has the fixed charge on spherical surfaces (described by $\sigma_i^\text{f}$, see Remark~\ref{free_surf_charge_remark}). Let us now focus on the first case, as the second one can be treated in a very similar way. For the mathematical convenience, we introduce the scaling of the coefficients of \eqref{Lin_eqs_Phi_in} for any $i$~as 
\begin{equation}
\label{L_nm_scaling}
L_{n m,i} \mathrel{:=} \frac{\Tilde L_{n m,i}}{(2 n+1) a_i \Tilde a_i^n} ,
\end{equation}
so that the boundary potential $\left.\Tilde\Phi_{\text{in},i}\right|_{r_i=a_i} = \sum\nolimits_{n,m}\Tilde L_{n m,i} \frac{Y_n^m(\Hat{\mathbf r}_i)}{(2 n+1) a_i} \in H^1(\partial\Omega_i)$ and thence $\sum_{n,m}|\Tilde L_{n m,i}|^2 < \infty$ holds. Since energy expansion \eqref{energy_expansion_components_abs_gen} with addends \eqref{energy_expansion_components_abs_center} stems from the series~\textcolor{red}{\cite{supplem_pre}} 
\begin{equation}
\label{energy_expansion_general_center}
\mathcal E = \frac{1}{2}\sum_{i=1}^N\frac{\varepsilon_i \varepsilon_0}{\kappa}\sum_{n,m}(2 n+1)\Hat L_{n m,i}^\star L_{n m,i} 
\end{equation}
(which in turn follows form the energetic integral 
\begin{equation}
\label{energy_expansion_general_integral_center}
\mathcal E = \frac{1}{2}\sum_{i=1}^N\int_{\Omega_i} \rho_i^\text{f}(\mathbf r) \Tilde\Phi_{\text{in},i}(\mathbf r) d\mathbf r
\end{equation}
and the definition of moments \eqref{varPhi_in_i_multipoles_Hat_Lmn}, see \textcolor{red}{\cite{supplem_pre}} for details; the $\Hat\varPhi_{\text{in},i}$-potential contributions do not affect the convergence of the screening-ranged energy expansion and are therefore omitted in this proof), hence it is useful to first examine the absolute convergence of \eqref{energy_expansion_general_center}. Indeed, substituting \eqref{L_nm_scaling} we have~$\forall i$:
\begin{align}
& \sum_{n,m}|(2 n+1)\Hat L_{n m,i}^\star L_{n m,i}| =  a_i^{-1}\sum_{n,m}|\Tilde a_i^{-n} \Hat L_{n m,i}^\star \Tilde L_{n m,i}| \label{energy_series_general_convergence}\\
&\quad \left|\text{use the Cauchy-Schwarz inequality}\right| \notag\\
&\le a_i^{-1} \sqrt{\sum\nolimits_{n,m}|\Tilde a_i^{-n} \Hat L_{n m,i}^\star|^2} \sqrt{\sum\nolimits_{n,m}|\Tilde L_{n m,i}|^2} < \infty \notag
\end{align}
since the series $\sum_{n,m}|\Tilde a_i^{-n} \Hat L_{n m,i}^\star|^2$ converges -- indeed, using \eqref{varPhi_in_i_multipoles_Hat_Lmn}, estimating its integral modulus and employing the elementary inequality \cite[Eq.~(3.69)]{Jack} $|Y_n^m(\Hat{\mathbf r}_i)| \le \frac{\sqrt{2 n+1}}{\sqrt{4\pi}}$ 
we then have $\Bigl|\frac{\Hat L_{n m,i}^\star}{\Tilde a_i^n}\Bigr| \le \frac{\kappa M_i \mathfrak g_i^n}{\sqrt{(2 n+1)4\pi}\varepsilon_i\varepsilon_0}$ with $M_i\mathrel{:=} \int_{\Omega_i} |\rho_i^\text{f}(\mathbf r_i)| d\mathbf r_i <\infty$ and $\mathfrak g_i \mathrel{:=} \frac{r_{i,\text{max}}}{a_i} < 1$, where $r_{i,\text{max}}$ estimates the maximal radial distance $r_i$ of points $\mathbf r_i$ inside sphere $\Omega_i$ where density $\rho_i^\text{f}(\mathbf r_i)$ is non-vanishing ($r_{i,\text{max}}<a_i$ since $\Omega_i$ entirely contains the support of $\rho_i^\text{f}$, $\operatorname{supp}(\rho_i^\text{f})\subset\Omega_i$). Hence $\sum\limits_{n,m}\bigl|\frac{\Hat L_{n m,i}^\star}{\Tilde a_i^n}\bigr|^2 \le \frac{\kappa^2 M_i^2}{4\pi\varepsilon_i^2\varepsilon_0^2} \sum\limits_{n=0}^{+\infty} \mathfrak g_i^{2 n} < \infty $. The same arguments also ensure that the exchange of the signs of the infinite sums and the integral when deriving \eqref{energy_expansion_general_center} from \eqref{energy_expansion_general_integral_center} was indeed permissible (the corresponding series converges uniformly inside the sphere and can therefore be integrated term-by-term).

From \eqref{scr_ranged_expansion_L} and scaling \eqref{L_nm_scaling} we immediately rewrite the screening-ranged expansion \eqref{scr_ranged_expansion_L} as $\Tilde L_{n m,i} = \sum_{\ell=0}^{+\infty} \Tilde L_{n m,i}^{(\ell)}$ with components $\Tilde L_{n m,i}^{(\ell)} \mathrel{:=} (2 n+1) a_i \Tilde a_i^n L_{n m,i}^{(\ell)}$. Taking this into account we have from~\eqref{energy_expansion_general_center}:
\begin{subequations}
\begin{align}
\mathcal E & = \frac{1}{2}\sum_{i=1}^N\frac{\varepsilon_i \varepsilon_0}{\Tilde a_i}\sum_{n,m} \Tilde a_i^{-n} \Hat L_{n m,i}^\star \sum_{\ell=0}^{+\infty} \Tilde L_{n m,i}^{(\ell)} = \left|\text{see below}\right| \label{temp_energy_der_center_1}\\
& = \sum_{\ell=0}^{+\infty} \biggl(\frac{1}{2}\sum_{i=1}^N\frac{\varepsilon_i \varepsilon_0}{\Tilde a_i}\sum_{n,m} \Tilde a_i^{-n} \Hat L_{n m,i}^\star \Tilde L_{n m,i}^{(\ell)}\biggr) \mathrel{=:} \sum_{\ell=0}^{+\infty}\mathcal E^{(\ell)} \label{temp_energy_der_center_2}
\end{align}
\end{subequations}
and one then naturally gets
\begin{equation*}
|\mathcal E| \le \sum_{\ell=0}^{+\infty} |\mathcal E^{(\ell)}| \le \sum_{\ell=0}^{+\infty}\frac{1}{2}\sum_{i=1}^N\frac{\varepsilon_i \varepsilon_0}{\Tilde a_i}\sum_{n,m} \biggl|\frac{\Hat L_{n m,i}^\star}{\Tilde a_i^{n}} \Tilde L_{n m,i}^{(\ell)}\biggr| .
\end{equation*}
Hence, to prove the absolute convergence of the energy series \eqref{energy_expansion_components_abs_gen} with addends \eqref{energy_expansion_components_abs_center} (or, equivalently, \eqref{temp_energy_der_center_2} with the scaling \eqref{L_nm_scaling} applied), and also that the change in the orders of sums during the transition from \eqref{temp_energy_der_center_1} to \eqref{temp_energy_der_center_1} was indeed permitted, it is sufficient to prove the convergence of the multiple series $\sum_{\ell=0}^{+\infty} \sum_{n,m} \bigl|\Tilde a_i^{-n}\Hat L_{n m,i}^\star \Tilde L_{n m,i}^{(\ell)}\bigr|$. To do so we obtain~$\forall i$:
\begin{align*}
& \sum_{\ell=0}^{+\infty} \sum_{n,m} \bigl|\Tilde a_i^{-n}\Hat L_{n m,i}^\star \Tilde L_{n m,i}^{(\ell)}\bigr| \le \left|\text{use the Cauchy-Schwarz inequality}\right| \notag\\
&\quad \le \sqrt{\sum_{n,m}|\Tilde a_i^{-n} \Hat L_{n m,i}^\star|^2} \sum_{\ell=0}^{+\infty}\sqrt{\sum_{n,m}|\Tilde L_{n m,i}^{(\ell)}|^2} < \infty .\notag
\end{align*}
Indeed, $\sum_{n,m}|\Tilde a_i^{-n} \Hat L_{n m,i}^\star|^2 < \infty$ was proven above. As for the 2nd series, introducing the global block column-vector $\Vec{\Tilde{\mathbb L}}$ (of the same structure as $\Vec{\Tilde{\mathbb G}}$, see \eqref{global_lin_sys1}, but with the coefficients $\Tilde L_{n m,i}$ instead of $\Tilde G_{n m,i}$) and denoting the matrix representation of the above-introduced operator $\mathring{\mathcal M}$ by $\mathbb M$ (see \eqref{matrix_representation_M_for_L}), so that $\mathbb M\in B(\pmb l^2)$, the boundary condition \eqref{implement_bc1} can be realised as $\Vec{\Tilde{\mathbb L}}=\mathbb M \Vec{\Tilde{\mathbb G}} + \Vec{\text{RP}}$, where the Right Part block vector $\Vec{\text{RP}}$ consists of values $\frac{2 n+1}{-\kappa}\frac{\Hat L_{n m,i}^\star}{\Tilde a_i^{n}}$ and is square-summable (which immediately follows from the estimates above and the elementary series $\sum_{n=0}^{+\infty} (2 n+1)^2\mathfrak g_i^{2 n}<\infty$); plugging expansion \eqref{scr_ranged_expansion_G} into $\Vec{\Tilde{\mathbb L}}=\mathbb M \Vec{\Tilde{\mathbb G}} + \Vec{\text{RP}}$ the screening-ranged components $\Tilde L_{n m,i}^{(\ell)}$ can be recovered (see \eqref{scr_ranged_expansion_L_ell_explicit}) and screening-ranged block vectors $\Vec{\Tilde{\mathbb L}}^{(\ell)}$ can also be formed (similarly to the treatment of the components $\Vec{\Tilde{\mathbb G}}^{(\ell)}$ of \eqref{scr_ranged_expansion_G}). Now, $\sum_{\ell=0}^{+\infty}\sqrt{\sum_{n,m}|\Tilde L_{n m,i}^{(\ell)}|^2} \le \sum_{\ell=0}^{+\infty}\|\Vec{\Tilde{\mathbb L}}^{(\ell)}\|_{\pmb l^2}$ and the convergence easily follows from the boundedness of the block operator $\mathbb M$ and its blocks and estimates in~\eqref{temp_eq_G_ell_converge_l2}.

\section{Summary and outlook}
\label{summary_conclusions}
\noindent
In the current paper we analyze the spectra of the operators naturally arising in the scheme of eliminating the multipole coefficients of the spherical harmonics expansions of internal potentials in favor of those of external potentials, for the problem of many interacting dielectric spheres. This scheme leads to a highly nontrivial coupling between the potential expansion coefficients, and although similar approaches (and variations) have been repeatedly used in the chemical-physics and soft-matter literature, an appropriate rigorous infinite-dimensional spectral analysis has never been performed before. The main practical result of our spectral analysis is that it makes it possible to explicitly (without resorting to a numerical solution of systems of equations coupling potentials) express the solution of the many-body boundary value problem (with transmission type boundary conditions for matching potentials of different dielectric media) in the form of a specific operator series constructed in the work and whose convergence and properties are actually ensured by our analysis; this is the cornerstone of our formalism of screening-ranged expansions of electrostatic quantities elaborated in detail in other parts of this study \textcolor{red}{\cite{supplem_prl,supplem_pre,supplem_pre_force}}. To this end, we establish and analyze connections between the conventional ``discrete'' formulation (in the space of spherical Fourier coefficients of potentials) and the ``continuous'' (Boundary-Integral-Equations-type) nonstandard ones. We believe that this pathway may also be adapted to more advanced models (e.g.~including a Stern layer, which however leads to an additional set of boundary conditions and even more complicated structure of the underlying operators than those introduced in Sec.~\ref{basic_BIE_formulations}) as well as for problems with other types of boundary conditions (like fixed boundary potentials, linear charge regulation conditions, etc.).

\section{Acknowledgments}
\label{sec:Acknowledgements}

We acknowledge the financial support from the European 
Union - NextGenerationEU and the Ministry of University and Research (MUR), 
National Recovery and Resilience Plan (NRRP): 
Research program CN00000013 “National Centre for HPC, 
Big Data and Quantum Computing”, CUP: J33C22001180001, funded by the D.D. n.1031, 17.06.2022 and Mission 4, Component 2, Investment 1.4 - Avviso “Centri Nazionali” - D.D. n. 3138, 16 December 2021.

\section*{Data availability}
\noindent
Codes/data can be found in~\textcolor{red}{\cite{our_github_rep}}.

\appendix

\section{On the range of applicability of re-expansion~\eqref{Yu3_reexp}}
\label{Yu_reexpansion_restriction_removing}
\noindent
As it was already noted (see footnote~\ref{footnote_Yu3_range_validity}), the Hobson-type \cite{LianMa} identity \eqref{Yu3_reexp} was rigorously proven in \cite{Yu3} imposing the additional restriction $r_j>R_{i j}$. For the sake of completeness, let us now discuss that in fact this condition can be lifted. To this end, let us first assume that $\mathbf R_{i j}$ is directed along the $Z$-axis ($\mathbf R_{i j} = R\mathbf z$) and consider the function $k_0(\Tilde r_j)$ ($\propto$ the left-hand side of \eqref{Yu3_reexp} at $L=M=0$). Then it is easy to check directly that this function satisfies the PBE $\diamondsuit_{r_i} k_0(\Tilde r_j) = 0 $, where $\diamondsuit_{r_i} \mathrel{:=} \Delta_{r_i} - \kappa^2 I$ is the modified Helmholtz differential operator in spherical coordinates associated with the $i$-th sphere (with $\Delta_{r_i}$ denoting the Laplace operator with $r_i$ as the radial spherical coordinate, and $I$ is the identity operator), and $r_j$ is considered as depending on spherical coordinates of the $i$-th sphere parametrically with respect to fixed $\mathbf R_{i j}$ (with parameter $R$ in the current case) -- indeed, from the cosine law one immediately has $r_j = \sqrt{R^2 + r_i^2 - 2 R r_i \cos\theta_i} = r_j(r_i, \theta_i)$. When $r_i<R$ (which is always fulfilled in our situation, since we consider non-overlapping spheres and actually use \eqref{Yu3_reexp} only to impose boundary conditions), function $k_0(\Tilde r_j)$ is smooth and regular (finite) with respect to spherical coordinates of the $i$-th sphere and also including at its center ($r_i=0$), thus $k_0(\Tilde r_j)$ must have a univocal expansion in terms of elementary PBE solutions $\{i_{l_1}(\Tilde r_i)Y_{l_1}^{m_1}(\Hat{\mathbf r}_i)\}_{0\le|m_1|\le l_1}$ regular at $r_i=0$; the coefficients of such an expansion are independent of spherical coordinates of the $i$-th sphere (but should depend on $\mathbf R_{i j}$, or $R$ in the current case, since $r_j$ doing so) and are unique, therefore it suffices to know them at one (arbitrary) point in order to conclude that they serve for all points. However, comparing this now with the right-hand side of~\eqref{Yu3_reexp} one sees that the latter provides an expansion of exactly the form we need, with expansion coefficients determined by $\mathcal H_{l_1 m_1}^{L M}(\mathbf{R}_{i j})$ (with superscripts $L=M=0$ in the case under consideration); expansion~\eqref{Yu3_reexp} with these coefficients is therefore valid for \emph{all} points (provided that condition $r_i<R$ holds), not only for those with the additional condition $r_j>R_{i j}=R$ imposed.

However, it is not so straightforward to directly transfer the above approach to the case of arbitrarily oriented $\mathbf R_{i j}$ and general $(L,M)$ in~\eqref{Yu3_reexp}, as it leads to tedious and cumbersome calculations when checking out whether $\diamondsuit_{r_i} k_L(\Tilde r_j) Y_L^M(\Hat{\mathbf r}_j) = 0 $ holds. Thus, instead of direct calculations let us use the representation derived in \cite[Eq.~(2.11)]{CleSch} (see \cite[Sec.~II]{CleSch} for its tedious proof): $$k_L(\Tilde r_j) Y_L^M(\Hat{\mathbf r}_j) = (-1)^L Y_L^M \Bigl(\frac{1}{\kappa}\nabla_{r_j}\Bigr) k_0(\Tilde r_j) ,$$ where the differential operator $Y_L^M(\frac{1}{\kappa}\nabla_{r})$ is related with the spherical harmonic $Y_L^M(\theta, \varphi) = (e^{\imath\varphi} \sin\theta)^M\sum_{n=0}^{L-M}C_{n,L M}\cos^n\theta$ (where $C_{n,L M}$ are some numerical coefficients) through replacements $$\cos\theta\to\frac{1}{\kappa}\frac{\partial}{\partial z},\qquad e^{\imath\varphi} \sin\theta \to \frac{1}{\kappa}\Bigl(\frac{\partial}{\partial x}+\imath\frac{\partial}{\partial y}\Bigr);$$ note that $Y_L^{-M}(\theta,\varphi) = (-1)^M Y_L^M(\theta,\varphi)^\star$ and for negative azimuthal index $M$ one should use $e^{-\imath\varphi} \sin\theta \to \frac{1}{\kappa}\Bigl(\frac{\partial}{\partial x}-\imath\frac{\partial}{\partial y}\Bigr)$. Let us also note the well-known fact that Laplace operator is invariant under the action of the group of Euclidian motions (rigid-motion transformations (translations/rotations)) \cite{Jack} and that the spherical radial variable as well as functions depending solely on it (such as $k_0$) remain unchanged during rotations. Then, as $r_i<R_{i j}$ and vector $\mathbf R_{i j}$ is constant, one gets the commuting identity for differential operators acting on the analyticity domains of the functions under consideration, $$\diamondsuit_{r_i} k_L(\Tilde r_j) Y_L^M(\Hat{\mathbf r}_j) = (-1)^L Y_L^M \Bigl(\frac{1}{\kappa}\nabla_{r_j} \Bigr) \diamondsuit_{r_i} k_0(\Tilde r_j),$$ and now rotating synchronously coordinate systems so that $\mathbf R_{i j}$ is on the $Z$-axis we can reduce the situation to the case considered above (where it was demonstrated that $\diamondsuit_{r_i} k_0(\Tilde r_j) = 0$). Hence we can finally obtain $\diamondsuit_{r_i} k_L(\Tilde r_j) Y_L^M(\Hat{\mathbf r}_j) = 0$, furthermore $k_L(\Tilde r_j) Y_L^M(\Hat{\mathbf r}_j)$ is regular at the $i$-th sphere's center ($r_i=0$), and thereby all the previous argumentation concerning the expansion of $k_L(\Tilde r_j) Y_L^M(\Hat{\mathbf r}_j)$ over family $\{i_{l_1}(\Tilde r_i)Y_{l_1}^{m_1}(\Hat{\mathbf r}_i)\}_{0\le|m_1|\le l_1}$ now steps~in.~\qed

\section{A brief summary of useful facts on modified Bessel functions and spherical~harmonics}
\label{appendix_bessel_functions_summary}
\noindent
Here we collect some elementary facts on Bessel functions for the convenience of referencing throughout the text. Spherical modified Bessel functions of the first and second kinds, respectively $i_n(x)$ and $k_n(x)$ (with $n\ge0$),~are
\begin{equation}
\label{i_n_k_n_modified_Bessel}
i_n(x) \mathrel{:=} \sqrt{\frac{\pi}{2}}\frac{I_{n+1/2}(x)}{\sqrt{x}}, \qquad k_n(x) \mathrel{:=} \sqrt{\frac{2}{\pi}}\frac{K_{n+1/2}(x)}{\sqrt{x}},
\end{equation}
where modified Bessel functions of the first and second kinds of semi-integer order $n+1/2$, respectively $I_{n+1/2} ( \cdot )$ and $K_{n+1/2} ( \cdot )$, have exact analytic representations \cite{Wat, GradRyzh} $K_{n+1/2}(x)= \frac{\sqrt{\pi}}{\sqrt{2 x}} e^{-x}\sum_{l=0}^n\frac{(n+l)!}{l! (n-l)! (2 x)^l}$, $I_{n+1/2}(x) = \frac{1}{\sqrt{2\pi x}}\Bigl(e^x\sum_{l=0}^n\frac{(-1)^l (n+l)!}{l! (n-l)! (2 x)^l} + (-1)^{n+1} e^{-x} \sum_{l=0}^n \frac{(n+l)!}{l! (n-l)! (2 x)^l} \Bigr)$.

There are recurrences~\cite[Eq.~(8.486)]{GradRyzh} 
\begin{equation}
\label{in_kn_recurrences0}
\begin{aligned}
& x k_{n+1}(x) = x k_{n-1}(x) + (2 n+1)k_n(x), \\ 
& x i_{n+1}(x) = x i_{n-1}(x) - (2 n+1)i_n(x).
\end{aligned}
\end{equation}
In addition, for their derivatives there are the expressions
\begin{equation}
\label{diff_modifiedBessel}
\begin{aligned}
& \frac{d}{d x}i_n(x) = \frac{n}{x} i_n(x) + i_{n+1}(x), \\  
&\frac{d}{d x}k_n(x) = \frac{n}{x} k_n(x) - k_{n+1}(x).
\end{aligned}
\end{equation}

Functions $i_n(x)$ and $k_n(x)$ can also be expressed in terms of regular spherical Bessel and Hankel functions $j_n(x)$ and $h_n^{(1)}(x)$ as~$x>0$:
\begin{equation}
\label{relations_modified_and_regular_Bessel_Hankel}
i_n(x) = \imath^{-n} j_n(\imath x), \qquad k_n(x) = -\imath^{n} h_n^{(1)}(\imath x)
\end{equation}
($\imath$ is a complex unit). Next, for small $x\to0^+$ one has~\cite{Wat}
\begin{equation}
\label{small_Bessel_i_k}
\begin{aligned}
i_n(x) & = \frac{x^n}{(2 n+1)!!} + \frac{x^{n+2}}{2 (2 n+3)!!} + O(x^{n+4}) \\ 
& \sim \frac{x^n}{(2 n+1)!!}, \qquad  k_n(x) \sim \frac{(2 n-1)!!}{x^{n+1}},
\end{aligned}
\end{equation}
while for large $x\to+\infty$ one has $k_n(x) \sim \frac{e^{-x}}{x}$ (notation ``$f(x)\sim g(x)$ as $x\to y$'' here means that $f(x)$ behaves asymptotically like $g(x)$ as~$x\to y$). As well, for arbitrarily fixed $x>0$, one has~\cite{SidiHoggan} $I_\nu(x)\sim\frac{(x/2)^\nu}{\Gamma(1+\nu)}$ and $K_\nu(x)\sim\frac{1}{2}\frac{\Gamma(\nu)}{(x/2)^\nu}$ as $\nu\to+\infty$ ($\Gamma(\cdot)$ is the Euler Gamma function, $\Gamma(1+\nu)=\nu\Gamma(\nu)$),~thus 
\begin{equation}
\begin{gathered}
\label{mod_bessel_asympt_large_x}
i_n(x)\sim\frac{2^{n+1}(n+1)!\, x^n}{(2 n+2)!}, \quad k_n(x)\sim\frac{(2 n)!}{2^n n!\, x^{n+1}}, \\ i_n(x) k_n(x)\sim\frac{1}{(2 n+1) x}\quad\text{as}\quad n\to+\infty
\end{gathered}
\end{equation}
(note that the right-hand side asymptotics of $i_n(x)$ and $k_n(x)$ in \eqref{mod_bessel_asympt_large_x} in fact coincide with those of~\eqref{small_Bessel_i_k}).

Note also (see \cite{Wat}) that for arbitrary $x>0$ and $n\ge0$
\begin{equation}
\label{in_kn_are_positive_}
i_n(x)>0, \qquad k_n(x)>0.
\end{equation}

Finally, composing the Wronskian determinant $W[I_{n+1/2}(x), K_{n+1/2}(x)]$ (see \cite[Eq.~(8.474)]{GradRyzh}) we get a useful relation
\begin{equation}
\label{wronsky_in_kn}
i_n(x) k_{n+1}(x) + i_{n+1}(x) k_n(x) = x^{-2}.
\end{equation}

Spherical (complex-valued) harmonics are defined as~\cite{Jack} 
\begin{equation}
\label{Ynm_definition}
Y_n^m(\Hat{\mathbf r}_i) = \sqrt{\tfrac{(2 n+1)(n-m)!}{4\pi(n+m)!}} P_n^m(\cos\theta_i) e^{\imath m\varphi_i} ,
\end{equation}
where $P_n^m(x) = \frac{(-1)^m}{2^n n!}(1-x^2)^{m/2}\frac{d^{n+m}}{d x^{n+m}}(x^2-1)^n$ denotes the conventional associated Legendre polynomial and $\imath$ is a complex unit. 

\bibliography{LinPaper}

\end{document}